\documentclass[
    aps,
    prd,
    reprint,
    superscriptaddress, 
    nofootinbib
]{revtex4-2}

\usepackage{tikz}
\usepackage{amsmath}
\usetikzlibrary{shapes.geometric, arrows}
\usepackage{physics}
\usepackage{braket}
\usepackage{bm}
\usepackage{appendix}
\usepackage{float}
\usepackage{amssymb}
\usepackage{amsthm}
\usepackage{booktabs}

\usepackage[normalem]{ulem}
\usepackage{tabularx}
\usepackage{siunitx}
\usepackage{soul}

\usepackage{xcolor}

\newcommand{\gagg}{g_{a\gamma \gamma}}
\newcommand{\veff}{V_{\mathrm{eff}}}

\usepackage{comment} 

\begin{document}
\raggedbottom

\title{Photon-Counting Readout via Electro-Optic Microwave-to-Optical Transduction for Axion Haloscopes}

\author{Jing Wu}
\affiliation{Fermi National Accelerator Laboratory, Batavia, IL 60510, USA
}
\author{Raphael Cervantes}
\affiliation{Fermi National Accelerator Laboratory, Batavia, IL 60510, USA
}

\author{Do\~{g}a Murat K\"{u}rk\c{c}\"{u}o\~{g}lu}
\affiliation{Fermi National Accelerator Laboratory, Batavia, IL 60510, USA
}

\author{Zheshen Zhang}
\affiliation{Department of Electrical Engineering and Computer Science, University of Michigan, Ann Arbor, Michigan 48109, USA}

\author{Silvia Zorzetti}%
\affiliation{Fermi National Accelerator Laboratory, Batavia, IL 60510, USA
}\affiliation{Department of Physics and Astronomy, Northwestern University, Evanston, IL, 60208, USA}
\email{zorzetti@ieee.org}

\date{\today}
\begin{abstract}
Axion dark matter searches require highly sensitive detection schemes with high scan rate to efficiently explore the parameter space. Building upon the conventional Sikivie haloscope, where axions convert into microwave photons inside a magnetic-field-immersed resonant cavity, we present a photon-counting scheme based on an electro-optic microwave-to-optical transducer. Through a comprehensive theoretical model under realistic experimental constraints, we find that the optimal operating regime employs an overcoupled microwave cavity, trading peak transduction efficiency for reduced transducer noise. Furthermore, combining parameters already demonstrated in existing devices, in a device with optimized extrinsic coupling rates, would yield 
a scan-rate enhancement of $5$--$10$ relative to quantum-limited homodyne
detection. Improvements in quality factors, coupling rates, and thermal management push this enhancement beyond $50$.
While our quantitative model is benchmarked at $9\,\mathrm{GHz}$, the scheme adapts well to haloscopes across the $5\text{--}12\,\mathrm{GHz}$ range, where shrinking cavity volumes severely degrade the achievable scan rate. These results demonstrate the potential of electro-optical photon detection as a high-speed detection technique for axion dark matter searches.

\end{abstract}
\maketitle

\section{Introduction}
Dark matter is the non-luminous, non-absorbing matter that constitutes approximately 84.4\% of the universe’s matter and 26.4\% of its energy density~\cite{10.1093/ptep/ptac097}. Its gravitational effects are essential for understanding cosmic evolution and structure formation. The evidence for dark matter is overwhelming~\cite{Rubin:1982kyu, 10.1093/mnras/249.3.523, 1998gravitational_lensing, 10.1093/mnras/stw3385, Markevitch_2004, 2020Planck}, yet its fundamental nature remains unknown. Among the many proposed candidates, including weakly interacting massive particles (WIMPs), sterile neutrinos, axions, axion-like particles, and dark photons~\cite{10.1093/ptep/ptac097}, axions stand out as particularly compelling. Beyond being a viable dark matter candidate, axions address the long-standing strong CP problem in quantum chromodynamics (QCD): why does the neutron have an electric dipole moment close to zero, despite nonzero values being allowed in the QCD Lagrangian?~\cite{PhysRevLett.38.1440, PhysRevLett.40.223, PhysRevLett.40.279}.

Axions could be detected through their coupling to electromagnetism. A broad class of axion search experiments is designed to convert axions into photons within a strong static magnetic field. If this conversion happens inside a resonant cavity, whose resonant frequency $f_\mathrm{c}$ matches the axion mass through $hf_\mathrm{c} \approx m_ac^2$, the axion signal power can be resonantly enhanced known as a Sikivie haloscope~\cite{PhysRevLett.51.1415}. The axion-photon coupling is parametrized by $\gagg$ and is constrained by the solution to the strong CP problem. Because the axion mass is unknown, haloscopes must be tunable to search a wide range of frequencies. In current experiments, the signal power is expected to be $\SI{e-24}-\SI{e-22}{~W}$. Detecting such a weak signal requires a strong magnetic field, a high-$Q$ cavity in millikelvin temperatures, and low-noise quantum sensors. The scan rate is a crucial figure of merit and is related to various experimental parameters by
\begin{equation}\label{eqn:scan_rate}
R \propto B_0^4 \veff^2 Q_\mathrm{L} T_n ^{-2},
\end{equation}
where $B_0$ is the external magnetic field, $\veff$ is the effective detection volume, $Q_\mathrm{L}$ is the loaded quality factor, and $T_n$ is the system noise temperature. The Axion Dark Matter eXperiment (ADMX) uses a \si{136}{L} cavity operating at the TM$_{010}$ mode at 1 GHz to search for QCD axions with masses of a few \si{\mu eV} at a scan rate of \SI{543}{MHz/year}~\cite{PhysRevLett.120.151301}. However, higher masses lead to smaller volume cavities ($V\propto f^{-3}$), more metallic losses ($Q_L\propto f^{-2/3}$), and increased noise from the Standard Quantum Limit (SQL) ($T_n \propto f$), causing the scan rate to scale as $R \propto f^{-8.66}$. The same ADMX haloscope scaled to \SI{10}{GHz} would have a scan rate of $R = \SI{1.2}{Hz/year}$, which is untenably slow. For this reason, the QCD axion parameter space is largely unexplored above \SI{1}{GHz}. Yet many theoretical predictions place the axion mass above 2~GHz, strongly motivating the exploration of this parameter space. It is imperative to develop detector and quantum sensing technologies to make this parameter space accessible~\footnote{Increasing $\veff$ with more sophisticated resonator designs (MADMAX~\cite{c749-419q}, Orpheus~\cite{cervantes2022admx, Cervantes2022b}, ALPHA~\cite{PhysRevD.107.055013}, VERA~\cite{PhysRevApplied.21.L041002}, LAMPOST~\cite{PhysRevLett.128.231802}) is also an active area of research.}.

Conventional axion haloscopes rely predominantly on microwave amplifiers and homodyne-based readout schemes~\cite{PhysRevD.50.4744,PhysRevD.42.1297,PhysRevLett.80.2043,HAYSTACPhase1}. For a phase-preserving linear amplifier, the added quantum noise imposes a fundamental noise floor, corresponding to at least one input-referred noise photon~\cite{RevModPhys.82.1155,PhysRevD.26.1817,PhysRevB.83.134501,PhysRevLett.104.041301, PhysRevLett.118.061302,PhysRevLett.118.061302}. Although single-mode squeezing has been developed to enhance the scan rate beyond the quantum limit, its practical performance is constrained by finite squeezing and degradation of the squeezed state during amplification and transmission~\cite{PhysRevX.9.021023,Backes2021,HAYSTACPhase2}. Photon-counting detection provides a different route that can circumvent the added-noise constraint associated with amplification, since it measures photon number rather than field quadratures~\cite{PhysRevX.15.021031,PhysRevLett.126.141302}. In the microwave domain, however, photon-counting schemes are challenged by limited detection efficiency, bandwidth, and relatively high dark count rates~\cite{Inomata2016Single,Pankratov2022Towards,Pankratov2025Detection}. Microwave-to-optical quantum transduction offers an alternative approach that could potentially overcome these limitations by transferring the microwave photons to the optical regime~\cite{Lauk_2020,Holzgrafe:20,Sahu2022Quantum,Qiu2023Coherent,Wang2022}.
Once converted, the signal is read out using superconducting nanowire single-photon detectors (SNSPDs), which provide exceptional detection efficiency with an ultra-low dark count rate~\cite{natarajan2012superconducting, PhysRevLett.128.231802}.

In this work, we propose and analyze a readout scheme for axion haloscopes based on microwave-to-optical transduction followed by optical photon counting. We evaluate the scheme using an electro-optic transducer designed at $9~\mathrm{GHz}$ as a benchmark, although it can be adapted to other haloscope frequencies by redesigning the transducer's microwave resonator. Our theoretical model incorporates key experimental parameters of the transducer, including the single-photon coupling rate, the quality factors (intrinsic loss rates) , microwave and optical extrinsic coupling rates, pump power, and repetition rate. Furthermore, the model accounts for noise contributions from the sensing cavity, the transducer microwave mode, and detector dark counts. We investigate the trade-off between the transduction efficiency and total noise, and optimize the extrinsic coupling coefficients to maximize the scan-rate enhancement with respect to homodyne-detection base line. Under benchmark experimental conditions, we find that enhancements of $E\approx5-10$ are possible with parameters demonstrated in existing devices, operating with optimized extrinsic coupling rates. Furthermore, enhancements in excess of $50$ are accessible when pushing experimental conditions to their limits. These results highlight the potential of low-loss, low-noise, and high-repetition-rate
electro-optic transducers for rapid axion searches and provide a promising
pathway toward high-speed axion experiments in high frequencies.


\section{Axion signal model}
\label{sec:axion_signal}
In a haloscope setup, axions from the galactic dark matter halo are converted into microwave photons inside a resonant cavity permeated by a strong static magnetic field $\mathbf{B}_0$, via the inverse Primakoff effect. In natural units, the expected signal power $P$ is given by~\cite{PhysRevLett.104.041301}:
\begin{equation}
P = g_{a\gamma\gamma}^2 \frac{\rho_\mathrm{a}}{m_\mathrm{a}} B_0^2 V_{\mathrm{eff}} Q_L \frac{\beta}{\beta + 1} \mathcal{L}(\Delta),
\label{eq:axion-power}
\end{equation}
where $g_{a\gamma\gamma}$ is the axion-photon coupling constant, $\rho_\mathrm{a} \approx 0.45\text{ GeV/cm}^3$ is the local dark matter density, $m_\mathrm{a}$ is the axion mass, $V_{\mathrm{eff}}$ is the effective cavity volume including the geometric form factor, $Q_L$ is the loaded quality factor of the cavity and $\beta=\kappa_\mathrm{ext}/\kappa_\mathrm{i}$ is the coupling coefficient of sensing cavity. Here, $\kappa_{\mathrm{ext}}$ is the extrinsic coupling rate, $\kappa_{\mathrm{i}}$ is the intrinsic loss rate and $\mathcal{L}(\Delta)$ is the Lorentzian lineshape function given by
\begin{equation}
    \mathcal{L}(\Delta) = \frac{1}{1 + 4\Delta^2},
\end{equation}
where the dimensionless frequency detuning is defined as $\Delta = (f_\mathrm{a} - f_\mathrm{c}) / \Delta f_\mathrm{c}$. Here, $f_\mathrm{a}$ and $f_\mathrm{c}$ denote the axion frequency and the cavity's resonance frequency, respectively, while $\Delta f_\mathrm{c}$ represents the cavity's bandwidth. Due to the virialized velocity dispersion of dark matter in the galactic halo, the axion signal exhibits a characteristic Maxwell-Boltzmann-like lineshape centered at frequency $f_\mathrm{a} = m_\mathrm{a} c^2 / h$, with a relative bandwidth of $\Delta f_\mathrm{a} / f_\mathrm{a} \sim 10^{-6}$ (corresponding to a quality factor of $Q_\mathrm{a} \equiv f_\mathrm{a} / \Delta f_\mathrm{a} = 10^6$)~\cite{Sikivie}. 
To incorporate this weak signal into our quantum input-output framework, we describe the axion as a coherent drive in sensing cavity with annihilation operator $\hat{b}(t)$. The intracavity field operator $\hat{a}(t)$ evolves according to~\cite{GardinerInputOutput,RevModPhys.82.1155}
\begin{equation}
    \frac{d\hat{a}}{dt} = -i \omega_\mathrm{c} \hat{a} - \frac{\kappa_{\mathrm{total}}}{2} \hat{a} - \sqrt{\kappa_{\mathrm{ext}}}\hat{a}_{\mathrm{in}} - \sqrt{\kappa_\mathrm{i}}\hat{a}_{\mathrm{i}} - i \kappa_a \hat{b},
\end{equation}
where $\omega_\mathrm{c} = 2\pi f_\mathrm{c}$ is the angular resonance frequency of the cavity, $\kappa_{\mathrm{total}} = \kappa_{\mathrm{ext}} + \kappa_{\mathrm{i}}$ represents the total cavity decay rate, and $\kappa_\mathrm{a}$ is the axion-photon coupling rate. The operators $\hat{a}_{\mathrm{in}}(t)$ and $\hat{a}_{\mathrm{i}}(t)$ denote the input vacuum and noise field from the transmission line and the intrinsic bath noise, respectively. 
Solving for the steady state (see Appendix~\ref{apd:Axion-state}), we obtain the output state from the sensing cavity as a Gaussian state with mean vector and covariance matrix given by
\begin{subequations}
\begin{align}
    \label{eq:sensing-output}
    \boldsymbol{d}_\mathrm{out} &= \sqrt{2n_\mathrm{a}} \begin{pmatrix} \cos\phi_{\mathrm{a}} &\sin\phi_\mathrm{a} \end{pmatrix}, \\
    \boldsymbol{V}_{\mathrm{out}} &= \left( \frac{1}{2} + \frac{4\beta}{(1+\beta)^2}\mathcal{L}(\Delta)n_\mathrm{s} \right) \boldsymbol{I},
\end{align}
\end{subequations}
where $n_\mathrm{a}=P/(\hbar\omega_\mathrm{a}\Delta f_\mathrm{a})$ is the photon number of axion, $\phi_\mathrm{a}$ is the phase of the axion-induced microwave field and $n_\mathrm{s}$ is the thermal photon number of sensing cavity.

\section{Transducer-Based Axion Sensing}
In this section, we present a comprehensive theoretical analysis of axion readout scheme based on the microwave-to-optical transduction and benchmark its performance with microwave homodyne detection. We trace the signal and noise propagation across the conversion and photon detection, and evaluate the resulting scan rate and the scan-rate enhancement relative to homodyne detection.
\begin{figure}[bt]
    \centering
    \includegraphics[width=0.90\linewidth]{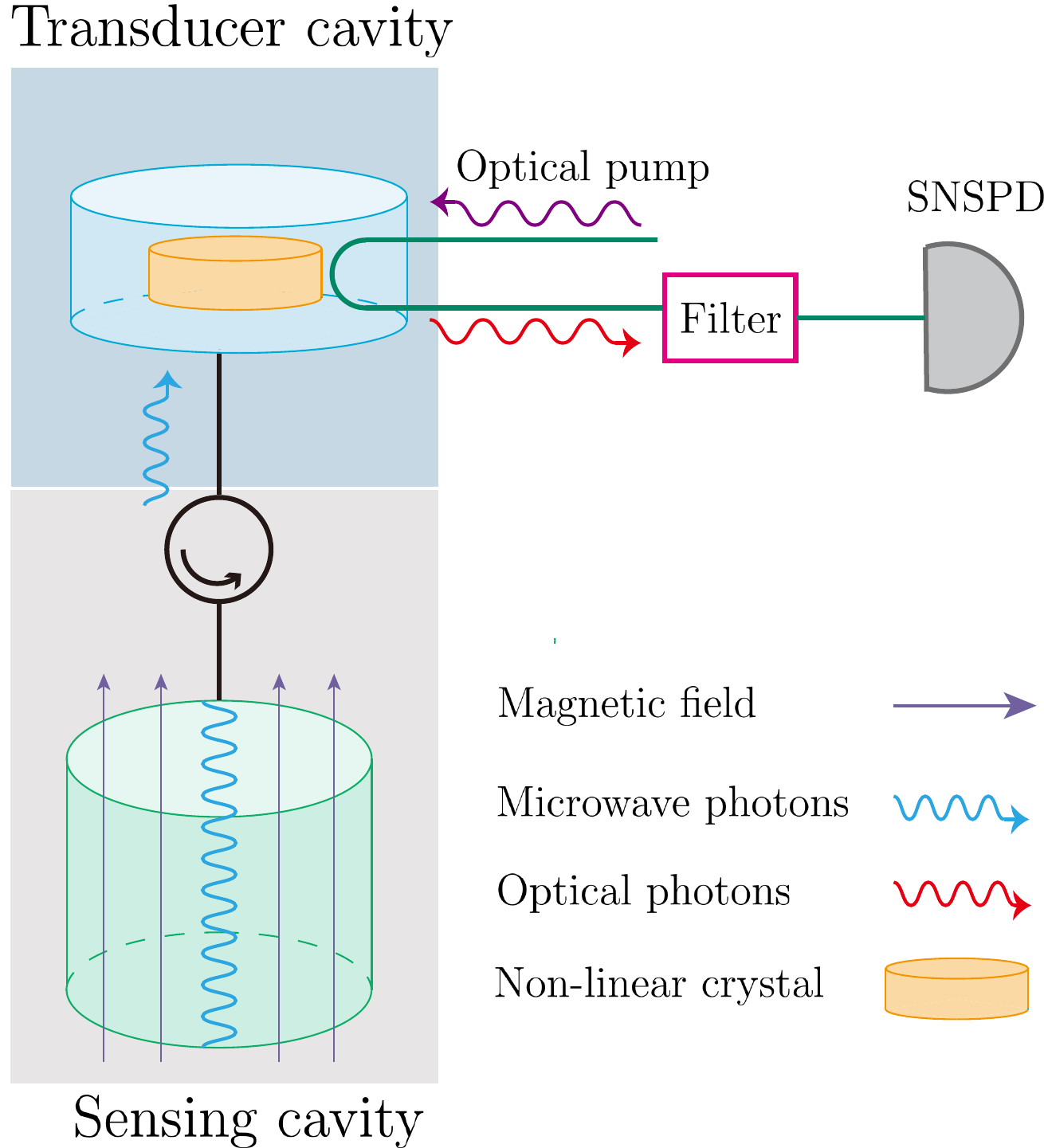}
    \caption{Schematic of the transducer-based axion sensing scheme. A strong magnetic field in the sensing cavity generates axion-induced microwave photons, which are routed to a transducer cavity and converted into optical photons through an optical pump and a nonlinear crystal for detection with a superconducting nanowire single-photon detector (SNSPD).}
    \label{fig:scheme}
    \vspace{-5 mm}
\end{figure}
\subsection{Signal and Noise Transfer}
Assuming a steady-state conversion process, the optical output field operator $\hat{a}_\mathrm{out}$ can be obtained from the transducer's dynamical equations, together with the corresponding input-output relations~\cite{Cavityquantumelectro-optics,Cavityoptomechanics}:
\begin{subequations}
\begin{align}
    &0= i\Delta_\mathrm{o} \hat{a} - ig\hat{m} - \frac{\gamma_\mathrm{o}}{2}\hat{a} 
    + \sqrt{\gamma_{\mathrm{oe}}}\hat{a}_{\mathrm{in}} + \sqrt{\gamma_\mathrm{{oi}}}\hat{a}_{\mathrm{i}}, \\
    &0= i\Delta_\mathrm{m} \hat{m} - ig\hat{a} - \frac{\gamma_\mathrm{m}}{2}\hat{m} 
    + \sqrt{\gamma_{\mathrm{me}}}\hat{m}_{\mathrm{in}} + \sqrt{\gamma_\mathrm{{mi}}}\hat{m}_{\mathrm{i}}, \\
    &\hat{a}_{\mathrm{out}} = -\sqrt{\gamma_{\mathrm{oe}}}\hat{a} + \hat{a}_{\mathrm{in}},\\
    &\hat{m}_{\mathrm{out}}= -\sqrt{\gamma_{\mathrm{me}}}\hat{m} + \hat{m}_{\mathrm{in}},
\end{align}
\end{subequations}
Here, we adopt a systematic subscript convention: the primary index ($\mathrm{o}$ or $\mathrm{m}$) denotes the subsystem (optical or microwave), whereas the secondary index ($\mathrm{c}$ or $\mathrm{i}$) denotes the extrinsic coupling and intrinsic loss channels, as summarized in Table~\ref{tab:variables} of Appendix~\ref{apd:transducer}. $\Delta_\mathrm{o}$ and $\Delta_\mathrm{m}$ are the optical and microwave detunings with respect to the output signals. The total decay rate of each mode is the sum of these two contributions,
$\gamma_\mathrm{o}=\gamma_\mathrm{oe}+\gamma_\mathrm{oi}$ and
$\gamma_\mathrm{m}=\gamma_\mathrm{me}+\gamma_\mathrm{mi}$, and
$g=\sqrt{n_\mathrm{p}}\,g_\mathrm{eo}$ is the pump-enhanced coupling rate,
where $n_\mathrm{p}$ is the intracavity pump photon number and
$g_\mathrm{eo}$ is the single-photon electro-optic coupling rate. From the derivations presented in Appendix~\ref{apd:transducer}, the microwave-to-optical conversion efficiency $\eta_\mathrm{T}$ and the associated added noise $n_\mathrm{add}$ are given by
\begin{align}
    \eta_\mathrm{T}&=\zeta_{\mathrm{o}}\zeta_\mathrm{m} \frac{4C}{(1+C)^2} , \label{eq:transducer-eta}\\
    n_\mathrm{add}&=\zeta_{\mathrm{o}}(1-\zeta_\mathrm{m}) \frac{4C}{(1+C)^2} n_\mathrm{m},
\end{align}
where $C = 4g^2/(\gamma_\mathrm{o}\gamma_\mathrm{m})$ is the cooperativity, $\zeta_\mathrm{o} = \gamma_\mathrm{oe}/\gamma_\mathrm{o}$ and $\zeta_\mathrm{m} = \gamma_\mathrm{me}/\gamma_\mathrm{m}$ are extraction efficiencies of the optical and microwave modes, respectively, and $n_\mathrm{m}$ denotes the effective thermal occupancy of the transducer's microwave mode. The quantum state by quantum transduction is given by
\begin{align}
    \boldsymbol{d}_\mathrm{T}&= \sqrt{2\eta_{\mathrm{D}}\eta_\mathrm{T} n_\mathrm{a}}(\cos\phi_{\mathrm{a}},\sin\phi_\mathrm{a}),\\
    \boldsymbol{V}_\mathrm{T}&=\left(\frac{1}{2}+n_\mathrm{T}\right)\boldsymbol{I},
\end{align}
where the transducer noise photon $n_\mathrm{T}$ is given by
\begin{equation}
    n_\mathrm{T}=\zeta_{\mathrm{o}} \frac{4C}{(1+C)^2} \left[(1-\zeta_\mathrm{m})n_{\mathrm{m}}+\zeta_\mathrm{m}\frac{4\beta}{(1+\beta)^2}n_\mathrm{s}\right].
    \label{eq:transducer-n_T}
\end{equation}

\subsection{Scan-Rate Enhancement Relative to Homodyne Detection}
The ultimate performance of a dark matter haloscope experiment is governed by its capability to scan a broad frequency range for axion masses. Having established the signal and noise transfer of the transducer-based readout chain, we now formulate the signal-to-noise ratio (SNR), the scan rate, and its performance enhancement relative to microwave homodyne detection. 

A detection event occurs when one or more photons are registered on the SNSPD. For a quantum transducer operating at a photon-counting measurement rate $f_\mathrm{r}$, a total of $N = t_\mathrm{int} f_\mathrm{r}$ individual photon detection cycles are performed over an integration time $t_\mathrm{int}$. For a pulsed pump, $f_\mathrm{r}$ is the repetition rate, whereas for a continuous pump, $f_\mathrm{r} = \Delta f_\mathrm{a}$. As presented in Appendix~\ref{apd:scan-rate-photon-detection}, the SNR is given by
\begin{equation}
    \text{SNR} = \frac{\eta_\mathrm{D} \eta_\mathrm{T} n_\mathrm{a}}{\sqrt{ \eta_\mathrm{D} n_\mathrm{T}+n_\mathrm{D} }}\sqrt{t_\mathrm{int} f_r},
\end{equation}
where $\eta_\mathrm{D}$ is the efficiency of the single photon detector, $n_\mathrm{D}$ is the dark count photon number per detection window. We next consider a frequency scan over a total range $\Delta f_\mathrm{scan}$, discretized into $N_2$ frequency steps with a step size $\delta<\Delta f_\mathrm{c}$. Following the derivation detailed in Appendix~\ref{apd:scan-rate-photon-detection}, the resulting scan rate $R_\mathrm{T}$ required to achieve a target total signal-to-noise ratio $\mathrm{SNR}_\mathrm{total}$ is given by
\begin{equation}
    R_\mathrm{T} = \frac{\pi}{4\text{SNR}_\mathrm{total}^2}\frac{\eta_\mathrm{D}^2\eta_\mathrm{T}^2n_\mathrm{a,0}^2}{\eta_\mathrm{D} n_\mathrm{T}+n_\mathrm{D} }\Delta f_\mathrm{c} f_\mathrm{r},
    \label{eq:RT}
\end{equation}
where $n_\mathrm{a,0}$ is the peak axion-induced photon number $n_\mathrm{a}(\Delta=0)$. For conventional microwave homodyne detection, the SNR and scan rate are derived in Appendix~\ref{apd:scan-rate-homodyne-detection} as:
\begin{align}
    \text{SNR} &= \frac{n_\mathrm{a}}{\sqrt{2}n_\mathrm{in,equiv}}\sqrt{t_\mathrm{int}\Delta \nu}, \\
    R_\mathrm{H} &= \frac{\pi}{8\text{SNR}_\mathrm{total}^2}\frac{n_\mathrm{a,0}^2}{n_\mathrm{peak}^{3/2}n_\mathrm{base}^{1/2}}\Delta f_\mathrm{c} \Delta \nu \label{eq:RH},
\end{align}
where $\Delta \nu$ is the resolution bandwidth of the spectrum analyzer, and $n_\mathrm{in,equiv}$ is the input-equivalent noise photon number of  system given by Friis formula (See Eq.~\eqref{eq:apd-noise-transfer} of Appendix \ref{sec:amplifier_noise})
\begin{align}
    n_{\mathrm{in,equiv}} &= \underbrace{1 + n_{\mathrm{JPA}} + \frac{n_{\mathrm{HEMT}}}{G_{\mathrm{JPA}}} + \dots}_{n_{\mathrm{base}}} + \frac{4\beta}{(1+\beta)^2} n_{\mathrm{s}} \mathcal{L}(\Delta) \nonumber \\
    &\equiv n_{\mathrm{base}} + (n_{\mathrm{peak}} - n_{\mathrm{base}}) \mathcal{L}(\Delta),
    \label{eq:n-inequiv-def}
\end{align}
where $n_{\mathrm{peak}}$ and $n_{\mathrm{base}}$ are the peak and base noise photon numbers, respectively. Here, $n_\mathrm{JPA}$ and $n_\mathrm{HEMT}$ denote the noise contributions from the Josephson parametric amplifier (JPA) and the high electron mobility transistor (HEMT) amplifier, respectively, with $G_\mathrm{JPA}$ being the JPA gain. Combining Eq.~\eqref{eq:RT} and Eq.~\eqref{eq:RH}, the enhancement $E$ in scan rate with the same total SNR is given by:
\begin{equation}
    E \equiv \frac{R_\mathrm{T}}{R_\mathrm{H}}=\frac{2\eta^2_\mathrm{D}\eta_\mathrm{T}^2n_\mathrm{peak}^{3/2}n_\mathrm{base}^{1/2}}{\eta_\mathrm{D} n_\mathrm{T}+n_\mathrm{D} }\frac{f_\mathrm{r}}{\Delta\nu}.
    \label{eq:Enhancement}
\end{equation}

\section{Performance Analysis}
\label{sec:Performance}

Evaluating the performance enhancement requires knowledge of both the transducer efficiency and noise, both of which depend on the cooperativity $C$. In this section, we first express the cooperativity in terms of the intrinsic cooperativity $C_0$ and the tunable coupling coefficients $(\beta_\mathrm{o}, \beta_\mathrm{m})$. Then we summarize the transducer's device and noise parameters at $9~\mathrm{GHz}$ across a realistic range of values in Table~\ref{tab:parameters}. After setting up the benchmark parameters, we maximize the enhancement by optimizing the tunable coupling coefficients $(\beta_\mathrm{o}, \beta_\mathrm{m})$ and characterize the trade-off between the transducer efficiency and the noise governed by the microwave coupling coefficient $\beta_\mathrm{m}$. Furthermore, we explore the scan-rate enhancement of transducer-based photon counting over conventional homodyne detection across key operational parameters, including the single-photon coupling rate $g_\mathrm{eo}$, the microwave quality factor $Q_\mathrm{m}$, the pump power $P_\mathrm{pump}$, and the thermal noise $T_\mathrm{m}$ ($n_\mathrm{m}$).  
Our analysis considers two operating schemes of quantum transducer: continuous-wave pumping, which provides low added noise but is limited by the transduction efficiency, and pulsed pumping, which offers higher efficiency at the cost of increased noise and a lower repetition rate.

\subsection{Parametric Characterization of Cooperativity}
The overall system performance is dictated by the conversion efficiency $\eta_\mathrm{T}$ and noise $n_\mathrm{T}$. A critical engineering trade-off arises from the dual role of the extrinsic coupling rates $(\gamma_\mathrm{oe},\gamma_\mathrm{me})$. High extrinsic coupling rates are essential extracting the converted signals. On the other hand, increasing the extrinsic coupling also increases the total damping rate $(\gamma_\mathrm{o},\gamma_\mathrm{m})$, which in turn degrades the cooperativity since $C \propto 1/(\gamma_{\mathrm{o}}\gamma_\mathrm{m})$. We parameterize the cooperativity $C$ using the tunable coupling coefficients $(\beta_\mathrm{o}, \beta_\mathrm{m})$ and the intrinsic cooperativity $C_0$, which depends on the intra-cavity pump photon number $n_\mathrm{p}$, single-photon coupling rate and intrinsic loss rates. Using standard input-output theory, $n_\mathrm{p}$ is directly related to the input pump power $P_\mathrm{p}$.

For electro-optic transducers, the intrinsic loss rates are set by the intrinsic quality factors of the two resonators, $\gamma_\mathrm{oi} = \omega_\mathrm{o}/Q_\mathrm{o}$ and $\gamma_\mathrm{mi} = \omega_\mathrm{m}/Q_\mathrm{m}$, where, $\omega_{o,m}$ and $Q_{o,m}$, are the operating angular frequencies and the intrinsic quality factor of the optical and microwave modes, respectively. These parameters are determined by material absorption, surface scattering, and fabrication quality, and are independent of the optical pump during normal operation, except in high-power regimes where pump-induced losses will arise. In contrast, the extrinsic coupling rates $(\gamma_\mathrm{oe}, \gamma_\mathrm{me})$ can be engineered and optimized, for instance, by designing the waveguide-cavity spacing or antenna geometry. We parameterize the tunable extrinsic coupling rates using the optical and microwave coupling coefficient defined by $\beta_\mathrm{o} \equiv \gamma_{\mathrm{oe}}/\gamma_{\mathrm{oi}}$ and $\beta_\mathrm{m} \equiv \gamma_{\mathrm{me}}/\gamma_{\mathrm{mi}}$. Consequently, the extraction efficiency is expressed as $\zeta_\mathrm{o} = \beta_\mathrm{o}/(1 + \beta_\mathrm{o})$. We define the intrinsic cooperativity
\begin{equation}
    C_0=\frac{4g_\mathrm{eo}^2n_\mathrm{p}}{\gamma_\mathrm{oi}\gamma_\mathrm{mi}}.
    \label{eq:Co}
\end{equation}
The electro-optic cooperativity is expressed as:
\begin{equation}
C = \frac{4  g_\mathrm{eo}^2n_\mathrm{p}}{\gamma_\mathrm{o} \gamma_\mathrm{m}}=\frac{C_0}{(1+\beta_\mathrm{m})(1+\beta_\mathrm{o})}.
\end{equation}
Electro-optic transducers are commonly operated with pulsed optical pumps. The cooperativity $C$ is proportional to pump photon number $n_\mathrm{p}$, whereas heating and thus noise $n_\mathrm{m}$ follows the average power. A low duty cycle therefore buys high cooperativity within fixed cooling budget, a strategy applied in bulk~\cite{Hease2020, Sahu2022Quantum, Qiu2023Coherent} and thin-film~\cite{Xu2021, Warner2025} lithium niobate as well as in aluminum nitride~\cite{Fu2021}. For example, in the high-cooperativity regime demonstrated in Ref.~\cite{Sahu2022Quantum}, the intrinsic cooperativity reaches $C_0 \approx 4$ for $C = 1$, with $\zeta_\mathrm{o} = 0.58$ and $\zeta_\mathrm{m} = 0.408$. Consequently, we consider a realistic range of intrinsic cooperativity $C_0 \in [1, 10]$ for our performance analysis.

Both in the steady-state regime (under continuous pumping) and the quasi-steady-state regime (where the pulse duration exceeds the cavity ring-up time), the intrinsic cooperativity $C_0$ can be further parameterized by the single-photon coupling rate $g_\mathrm{eo}$ and the pump power $P_\mathrm{pump}$. The single-photon coupling rate $g_\mathrm{eo}$ is determined by the field overlap between
the microwave mode and optical mode. Based on the steady-state input-output relation, for a given pump power $P_\mathrm{pump}$, the intra-cavity pump photon number is expressed as~\cite{Wang2022}
\begin{equation}
n_\mathrm{p} = \frac{\gamma_\mathrm{oe}}{\Delta_\mathrm{p}^2 + (\gamma_\mathrm{o}/2)^2}\frac{P_\mathrm{pump}}{\hbar \omega_\mathrm{p}},
\end{equation}
where $\Delta_\mathrm{p}$ is the pump detuning and $\omega_\mathrm{p}$ is the angular frequency of the pump field. Therefore, given the intrinsic loss rates $(\gamma_\mathrm{oi}, \gamma_\mathrm{mi})$, single-photon coupling rate $g_\mathrm{eo}$, pump power $P_\mathrm{pump}$, and their respective resonance frequencies, the resulting cooperativity $C$ depends on the tunable coupling coefficients $(\beta_\mathrm{o}, \beta_\mathrm{m})$ to be optimized.

\subsection{Enhancement Regime for Electro-Optic Transducers}
\label{sec:enhancement-regime}

To evaluate the practical performance of the transducer-based axion haloscope, we consider an axion search targeting a frequency around 9\,GHz. Table~\ref{tab:parameters} summarizes the benchmark parameters in this subsection.

The axion signal is evaluated at a benchmark frequency of $f_\mathrm{a} = 9.0\,\mathrm{GHz}$ with a bandwidth of $\Delta f_\mathrm{a} = f_\mathrm{a}/Q_\mathrm{a} = 9.0\,\mathrm{kHz}$. The effective temperature of the sensing cavity has been reported to be below $60\,\mathrm{mK}$ in Ref.~\cite{HAYSTACPhase2}. We adopt $T_\mathrm{s} = 50\,\mathrm{mK}$, within this range and consistent with the temperatures routinely achieved in dilution refrigerators hosting haloscope cavities. The thermal photon number in the sensing cavity is then determined by the Bose-Einstein distribution:
\begin{equation}
    n_\mathrm{s} = \frac{1}{e^{\frac{\hbar\omega_\mathrm{c}}{k_\mathrm{B} T_\mathrm{s}}} - 1}.
    \label{eq:Bose-Einstein-distribution}
\end{equation}
For conventional homodyne detection, the effective noise photon number for standard microwave homodyne detection is defined by 

\begin{equation}
    n_\mathrm{Hom} \equiv n_\mathrm{JPA} + \frac{n_\mathrm{HEMT}}{G_\mathrm{JPA}} + \cdots.
\end{equation}
The dominant contribution comes from the JPA. We evaluate $n_\mathrm{Hom}$ (i.e., $n_\mathrm{base}$ in Eq.~\eqref{eq:n-inequiv-def}) assuming a total effective noise temperature of $100~\mathrm{mK}$ for the microwave components. The frequency resolution $\Delta \nu$ is typically chosen to be $1/3$ to $1/5$ of the axion bandwidth $\Delta f_\mathrm{a} = 9.0\,\mathrm{kHz}$. We set $\Delta \nu = 3\,\mathrm{kHz}$ for homodyne detection.

\begin{table}[htbp]
    \caption{Benchmark parameters for the transducer-based axion haloscope. Values are chosen based on experimental capabilities for searches near $9\,\mathrm{GHz}$}
    \label{tab:parameters}
    \begin{ruledtabular}
        \begin{tabular}{lcc}
            \textbf{Parameter} & \textbf{Symbol} & \textbf{Value} \\
            \hline
            \multicolumn{3}{c}{\textit{Axion Sensing Cavity}} \\
            \hline
            Sensing resonance frequency & $\omega_\mathrm{c}/(2\pi)$ & $9.0\,\mathrm{GHz}$ \\
            Loaded quality factor & $Q_{L}$ & $ 10^5$ \\
            Axion quality factor & $Q_{\mathrm{a}}$ & $ 10^6$ \\
            extrinsic coupling ratio & $\beta$ & 2.0\\
            Cavity temperature  &$T_\mathrm{s}$ ($n_\mathrm{s}$) & $50\, \mathrm{mK}$\\
            \hline
            \multicolumn{3}{c}{\textit{Electro-Optic Transducer}} \\
            \hline
            Pump frequency & $\omega_\mathrm{p}/(2\pi)$ & $192.421\,\mathrm{THz}$ \\                 
            Optical resonance frequency & 
            $\omega_\mathrm{o}/(2\pi)$ & $192.43\,\mathrm{THz}$ \\
            Microwave resonance frequency & $\omega_\mathrm{m}/(2\pi)$ & $9.0\,\mathrm{GHz}$ \\
            Optical quality factor & $Q_\mathrm{o}$ & $10^7$ \\
            Microwave quality factor & $Q_\mathrm{m}$ & $10^3\text{--}10^5$ \\            
            Single-photon coupling rate & $g_\mathrm{eo}/(2\pi)$ & $40\text{--}140\;\mathrm{Hz}$  \\
            Pump detuning & $\Delta_\mathrm{p}/(2\pi)$ & $10\;\mathrm{MHz}$ \\
            \multicolumn{3}{c}{\textit{Continuous-wave operation}} \\
            Pump power & $P_\mathrm{pump}$ & $100\text{--}800\;\mu\mathrm{W}$ \\
            Transducer temperature& $T_\mathrm{m}$ ($n_\mathrm{m}$)  & $100\text{--}300\;\mathrm{mK}$ \\
            Photon counting measurment rate & $\Delta f_\mathrm{a}$ & $9.0~\mathrm{kHz}$\\  
            \multicolumn{3}{c}{\textit{Pulsed operation}} \\
            Pump power & $P_\mathrm{pump}$ & $50\text{--}300\;\mathrm{mW}$ \\
            Microwave noise photon & $n_\mathrm{m}$ & $0.01\text{--}0.1$\\
            Intrinsic cooperativity & $C_0$ & $1\text{--}10$\\            
            Repetition rate & $f_\mathrm{r}$ & $100~\mathrm{Hz}$\\           
            \hline
            \multicolumn{3}{c}{\textit{Single Photon Detector}} \\
            \hline
            Single-photon detection efficiency & $\eta_\mathrm{D}$ & $0.98$ \\
            Dark photon counting rate & $r_\mathrm{D}$ & $1\,\mathrm{Hz}$ \\
             \hline
            \multicolumn{3}{c}{\textit{Homodyne Detection}} \\
            \hline
            Frequency bin resolution &$\Delta\nu$& $3~\text{kHz}$ \\
            Total temperature &$T_\mathrm{Hom}$ ($n_\mathrm{Hom}$) & $100\;\mathrm{mK}$\\
        \end{tabular}
    \end{ruledtabular}
\end{table}

\subsubsection{Continuous-wave pump operation}
\label{subsection:CW-operation}
For a continuous-wave (CW) electro-optic transducer operating in steady-state conversion, the total number of modes is given by the time-bandwidth product $\Delta f_\mathrm{a} t_\mathrm{int}$ over an integration time $t_\mathrm{int}$. Thus the photon-counting measurement rate $f_\mathrm{r}=\Delta f_\mathrm{a}$. The average dark count per mode is given by $n_\mathrm{D} = 1/\Delta f_\mathrm{a}$ for a dark count rate of $r_\mathrm{D} = 1~\mathrm{Hz}$. A pump power in the order $100~\mu\mathrm{W}$ yields lower added noise (i.e., the effective temperature $T_\mathrm{m}$) but also a lower transduction efficiency $\eta_\mathrm{T}$ than pulsed operation~\cite{Hease2020,Xu2021,Warner2025}. 
Pump detuning ($\Delta_\mathrm{p}$) is essential to mitigate photothermal instabilities and optical bistability caused by high-power pump absorption. Accordingly, we set the pump detuning to $\Delta_\mathrm{p} = 10\,\mathrm{MHz}$ as a representative experimental benchmark~\cite{Wang2022}. The optical quality factor is set to $10^7$. The microwave quality factor is $10^3\text{--}10^4$ for on-chip resonators and projected $10^4\text{--}10^5$ for state-of-the-art 3D hybrid microwave cavities~\cite{Wang2022, zorzetti2023millikelvin, reineri2026microwave}. In steady-state conversion, the relation between $n_\mathrm{m}$ and $T_\mathrm{m}$ follows the Bose-Einstein distribution in Eq.~\eqref{eq:Bose-Einstein-distribution}. The steady-state input-output relation directly establishes the relation between the input pump power $P_\mathrm{pump}$ and the intra-cavity photon number, uniquely determining $C_0$ and $C$. In contrast, pulsed operation outside the quasi-steady-state regime requires solving the time-dependent intra-cavity dynamics or utilizing an effective $C_0$.

\begin{figure}[tb]
    \centering
    \includegraphics[width=0.85\linewidth]{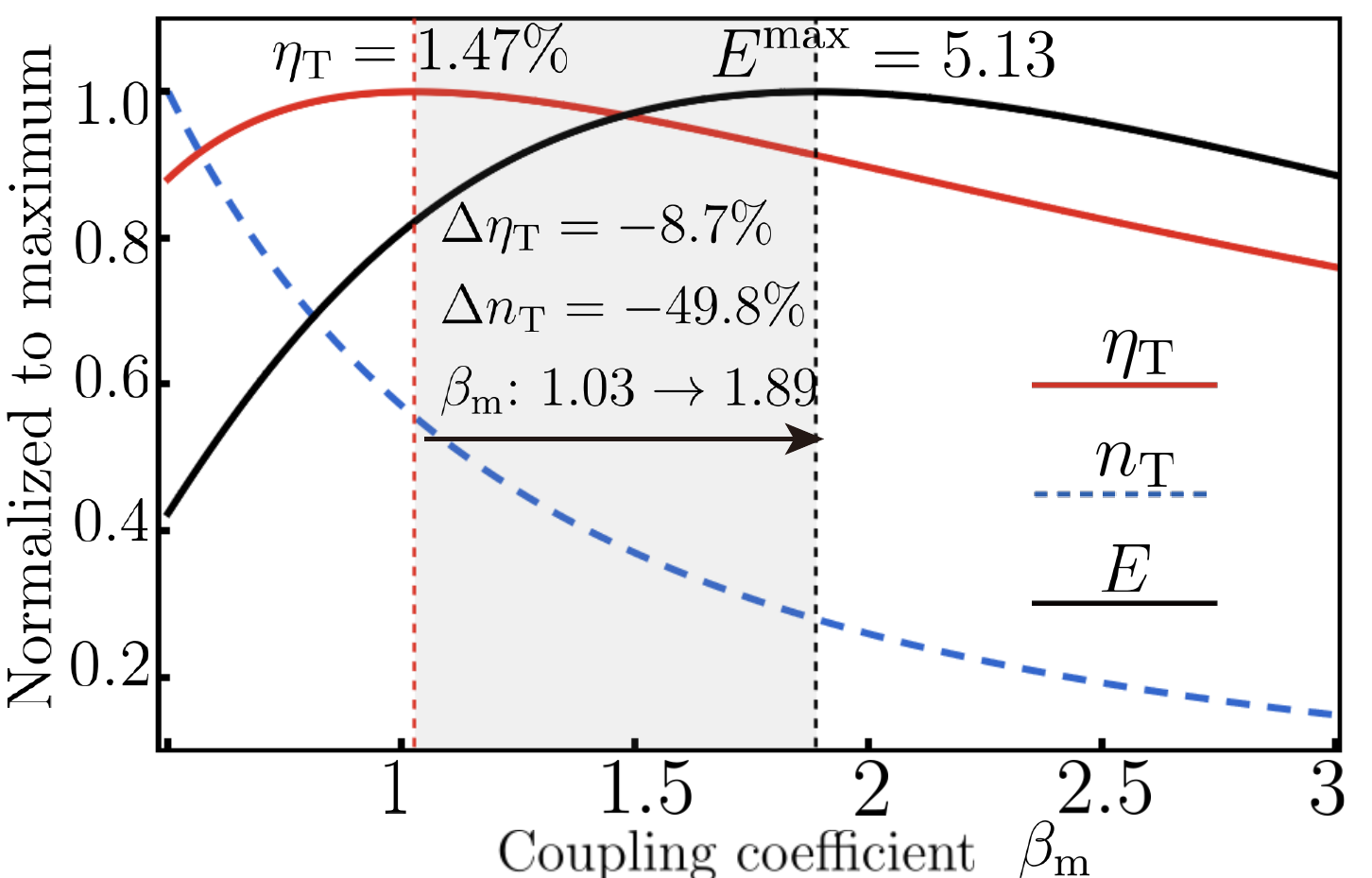}
\caption{Trade-off between transduction efficiency and noise in the
continuous-wave regime. Transduction efficiency $\eta_\mathrm{T}$ (red solid),
transducer noise $n_\mathrm{T}$ (blue dashed), and scan-rate enhancement $E$
(black solid) versus the microwave coupling ratio $\beta_\mathrm{m}$, with the
optical coupling fixed at its optimal value $\beta_\mathrm{o} = 1.23$.
Each curve is normalized to its own maximum
($\eta_\mathrm{T}^{\max} =1.47\%$, $n_\mathrm{T}^{\max} = 3.5\times10^{-4}$,
$E^{\max} = 5.13$). Vertical dashed lines mark the maxima of $\eta_\mathrm{T}$
($\beta_\mathrm{m} = 1.03$) and of $E$ ($\beta_\mathrm{m} = 1.89$); the shaded
band indicates the displacement between them. Here, $T_\mathrm{m} = 100\,\mathrm{mK}$ and $P_\mathrm{pump} = 400\,\mathrm{\mu W}$, $g_{eo}/(2\pi)=50~\mathrm{Hz}$, $Q_m=5\times10^4$.}
    \vspace{-3 mm}
    \label{fig:cw-trade-off}
\end{figure}

With $T_\mathrm{m}=100~\mathrm{mK}$, $g_\mathrm{eo}/(2\pi)=50~\mathrm{Hz}$, $Q_\mathrm{m}=5\times10^4$ and $P_\mathrm{pump}=400~\mu\mathrm{W}$, optimal choices of $(\beta_\mathrm{m},\beta_\mathrm{o})$ emerge from the trade-off between the overall transduction efficiency $\eta_\mathrm{T}$ and the noise $n_\mathrm{T}$. Through numerical simulations, we find a peak enhancement of $E = 5.13$ at $\beta_\mathrm{o} = 1.23$ and $\beta_\mathrm{m} = 1.89$, with a transducer efficiency of $\eta_\mathrm{T} = 1.34\%$. With $\beta_\mathrm{o}$ fixed at its optimum, we show the trade-off versus $\beta_\mathrm{m}$ in Fig.~\ref{fig:cw-trade-off}. Existing microwave-to-optical transducers were designed for quantum networking and qubit readout, where the goal is to maximize the conversion efficiency. They therefore operate near critical microwave coupling. The axion photon-counting scheme analyzed here instead favors a strongly overcoupled microwave port. We observe that a scan-rate enhancement is obtained at a relatively low transduction
efficiency. This is because photon counting is limited by noise photon $n_\mathrm{T}$ rather than by
the standard quantum limit ($n_\mathrm{in,equiv}\geq 1$), i.e., a weak signal can still be detected with a favorable signal-to-noise ratio. 

\begin{figure}[tb]
    \centering
\includegraphics[width=0.8\linewidth]{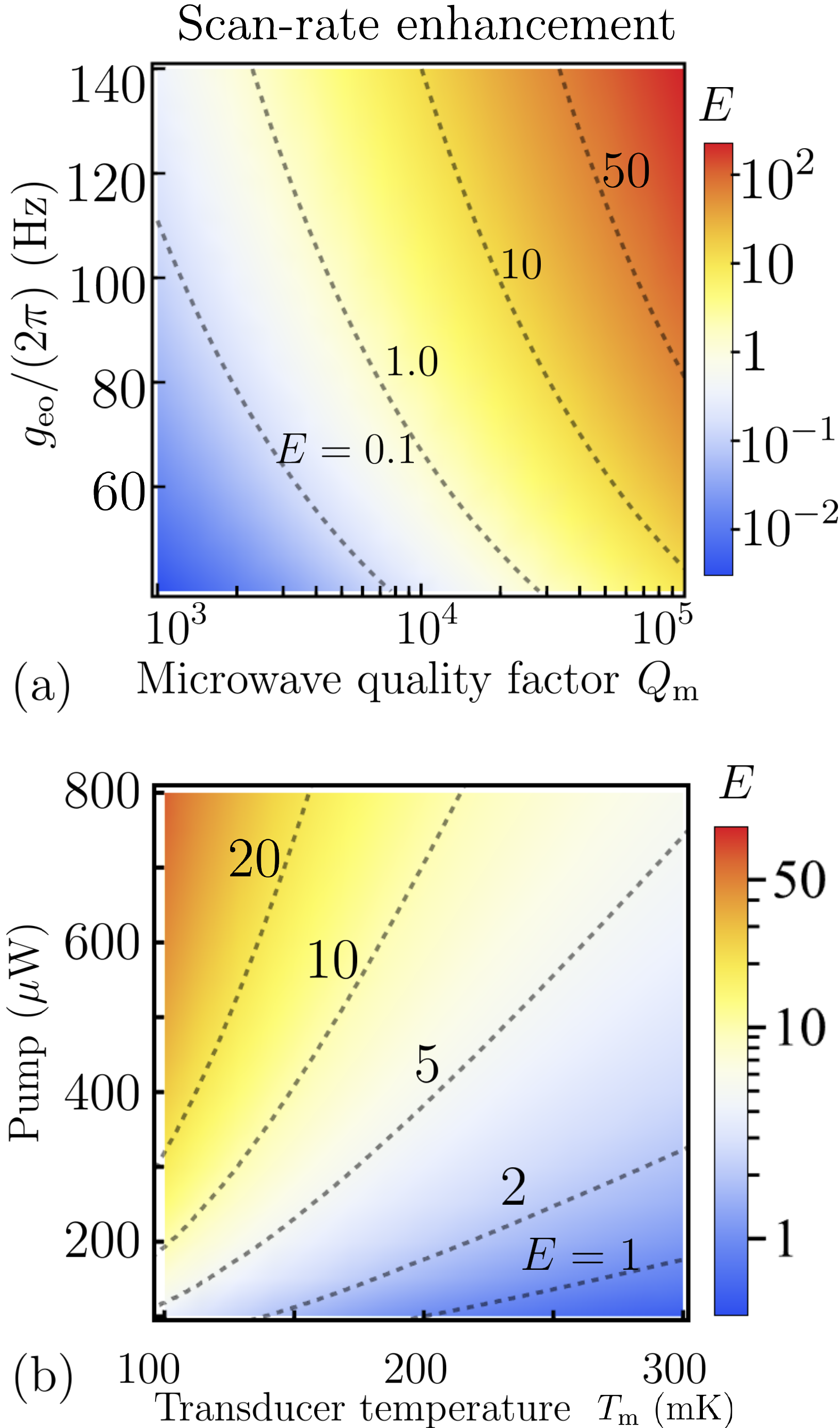}
    \caption{Scan-rate enhancement compared to standard homodyne detection under continuous-wave pumping. Dashed lines: the corresponding enhancement $E$. (a) Contour map of the optimal enhancement as a function of the $g_\mathrm{eo}$ and $Q_\mathrm{m}$, at a fixed temperature $T_\mathrm{m}=100~\mathrm{mK}$ and pump power $P_\mathrm{pump}=400~\mu\mathrm{W}$.  (b) Optimal enhancement versus $T_\mathrm{m}$ and $P_\mathrm{pump}$, at a fixed $Q_\mathrm{m}=5\times10^4$ and $g_\mathrm{eo}/(2\pi)=90~\mathrm{Hz}$. }
 \vspace{-3 mm}
    \label{fig:enhancementCW}
\end{figure}

Beyond these benchmark values, we evaluate the enhancement $E$, maximized over $(\beta_\mathrm{m}, \beta_\mathrm{o})$  for the parameters listed in Table~\ref{tab:parameters}. Fig.~\ref{fig:enhancementCW}(a) displays the enhancement over a range of single-photon coupling rate $g_\mathrm{eo}$ and microwave quality factor $Q_\mathrm{m}$, while Fig.~\ref{fig:enhancementCW}(b) shows the enhancement over a range of transducer temperature $T_\mathrm{m}$ and pump power $P_\mathrm{pump}$, with $E$ optimized over $(\beta_\mathrm{m}, \beta_\mathrm{o})$ at each point. The 3D microwave cavity transducer design in Ref.~\cite{Wang2022} features quality factors of $10^4\text{--}10^5$ with a relatively small single-photon coupling rate of $g_\mathrm{eo}/2\pi \approx 46\,\mathrm{Hz}$, placing its performance in the bottom-right regime of Fig.~\ref{fig:enhancementCW}(a) with $E = 11$. Conversely, on-chip thin-film lithium niobate transducers experiments in Ref.~\cite{Warner2025} offer a complementary route, exhibiting a much larger single-photon coupling rate ($g_\mathrm{eo}/2\pi \approx 945\,\mathrm{Hz}$) but lower quality factors ($10^2\text{--}10^3$). Consequently, such devices operate in the upper-left regime of Fig.~\ref{fig:enhancementCW}(a), yielding $E = 12$ at $Q_\mathrm{m} = 2.5 \times 10^2$. Although Fig.~\ref{fig:enhancementCW}(b) suggests that both a higher pump power and a lower transducer temperature increase the enhancement, these two parameters are not independent. A larger pump power raises the effective mode temperature through optical absorption and quasiparticle generation, with the precise dependence determined by the thermal anchoring of the device. In practice, current on-chip demonstrations are limited by fiber-to-chip losses and by the pump power they can sustain.

\subsubsection{Pulsed pump operation}
\label{subsection:Pulsed-operation}
While CW operation is constrained by steady-state heating, pulsed pumping allows high pump photon number for high cooperativity. Pulsed operation relaxes the thermal budget of the device: a peak pump power of $\sim100~ \mathrm{mW}$, applied at a duty cycle of $1–5\%$, dissipates only $\sim 1$ mW of average optical power at the mixing chamber while maintaining a high intra-cavity pump photon number $n_\mathrm{p}$, sufficient for larger cooperativity $C \geq 1$ and hence near unity internal efficiency $\eta_\mathrm{int}=4C/(1+C)^2$. Pump-induced heating means the mode occupancy relaxes only partially within one cycle and does not reach its steady-state ($\sim 100\mathrm{mK}$) value. In experiments, the thermal photon number $n_\mathrm{m}$ is typically on the order of $0.1$ at frequencies around $3\text{--}9~\mathrm{GHz}$~\cite{Hease2020, Sahu2022Quantum, Qiu2023Coherent, Xu2021, Warner2025, Fu2021, Sekine2026, werner2026electro}. Specifically, evaluated at $9~\mathrm{GHz}$, the effective input noise photon number falls within the range $n_\mathrm{m} = 0.01\text{--}0.1$. Since the bandwidth of transducer ($\sim1\text{--}10~\mathrm{MHz}$) is much larger than the signal bandwidth $\Delta f_\mathrm{a}$, the transducer responds essentially instantaneously. The pulse duration and detection time window is $1/\Delta f_\mathrm{a} \approx 111 \mu\mathrm{s}$, with dark photon number $n_\mathrm{D}=1/\Delta f_\mathrm{a}$ for $1~\mathrm{Hz}$ dark photon counting rate. 
For the numerical evaluation for pulsed pump operation, we fix the dark photon number $n_\mathrm{D}$ within this detection window and the repetition rate at $f_\mathrm{r} = 100\,\mathrm{Hz}$.
\begin{figure}[tb]
    \centering
    \includegraphics[width=0.85\linewidth]{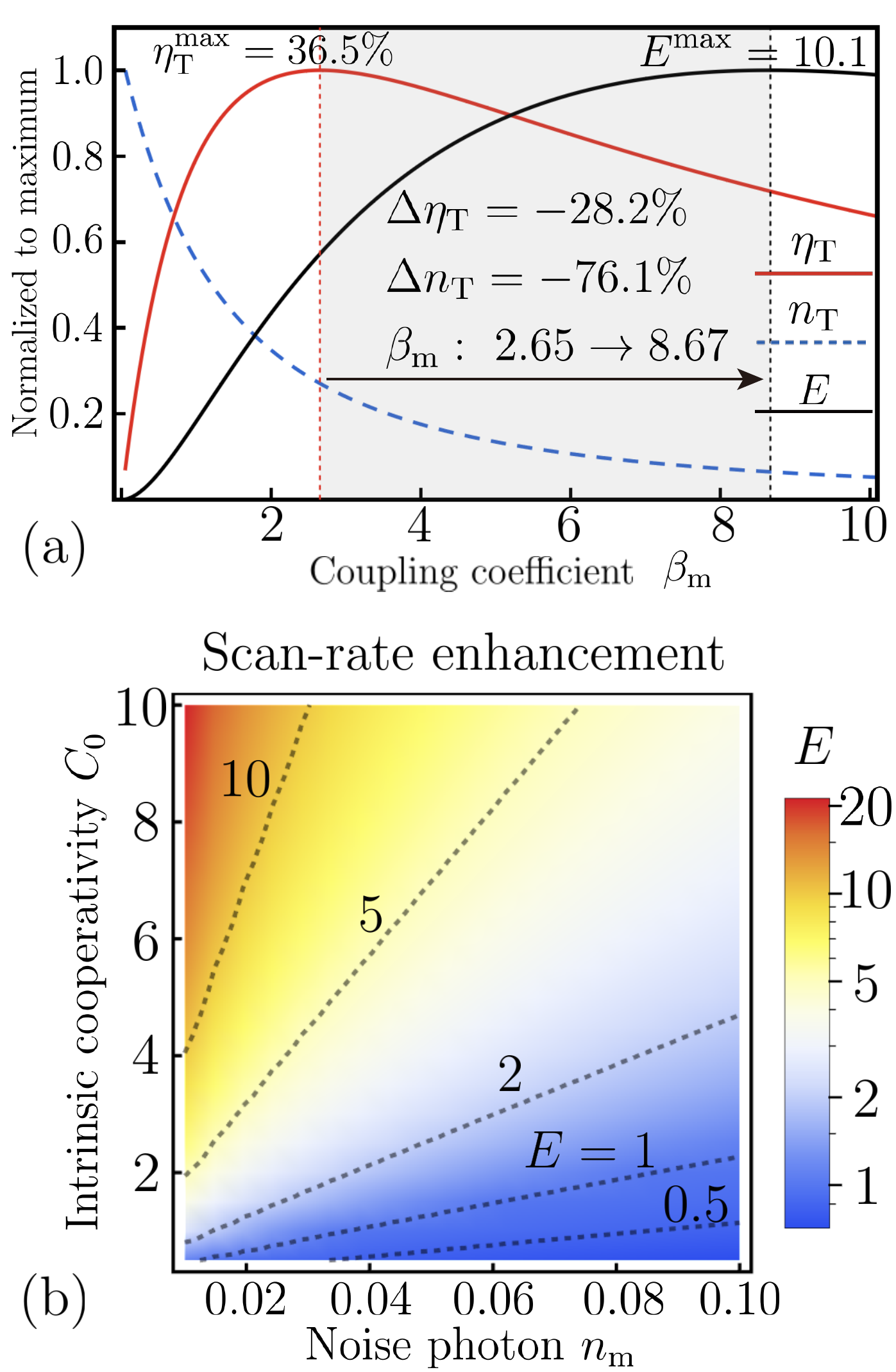}
\caption{(a) Trade-off between transduction efficiency and noise in the
pulsed-pump regime. Transduction efficiency $\eta_\mathrm{T}$ (red solid),
transducer noise $n_\mathrm{T}$ (blue dashed), and scan-rate enhancement $E$
(black solid) versus the microwave coupling ratio $\beta_\mathrm{m}$, with the
optical coupling fixed at its optimal value $\beta_\mathrm{o} = 1.41$.
Each curve is normalized to its own maximum
($\eta_\mathrm{T}^{\max} =36.5\%$, $n_\mathrm{T}^{\max} = 5.3\times10^{-3}$,
$E^{\max} = 10.1$). Here $C_0=4$ and
$n_\mathrm{m} = 0.01$. (b) Contour map of the enhancement as a function of the intrinsic cooperativity $C_0$ and transducer noise $n_\mathrm{m}$, at a fixed repetition rate $f_\mathrm{r}=100~\mathrm{Hz}$.}
    \vspace{-3 mm}
    \label{fig:pulse-trade-off}
\end{figure}

We now repeat the optimization for pulsed operation, using the parameters of
Table~\ref{tab:parameters}. With intrinsic cooperativity $C_0=4$ and noise photon number $n_\mathrm{m}=0.01$, Fig.~\ref{fig:pulse-trade-off}(a)
shows the enhancement as a function of $\beta_\mathrm{m}$.
We find a peak enhancement of $E = 10.1$ at $\beta_\mathrm{o} = 1.41$ and
$\beta_\mathrm{m} = 8.67$, for which the total transducer efficiency is
$\eta_\mathrm{T} =26.2\%$. The optimum again lies in the overcoupled regime. The large pump power lifts the internal efficiency, leaving more
efficiency to trade away; the optimal $\beta_\mathrm{m}$ therefore moves deeper
into the overcoupled regime, where $n_\mathrm{T}$ is further suppressed. Fig.~\ref{fig:pulse-trade-off}(b) shows the enhancement $E$ obtained by maximizing over the coupling ratios at each point of the $(n_\mathrm{m}, C_0)$ plane. The bulk-LiNbO$_3$ transducer of Ref.~\cite{Sahu2022Quantum} reaches cooperativities up to $C=0.92$, corresponding to $C_0\approx4$. For a repetition rate of $100\,\mathrm{Hz}$, the enhancement ranges from $E = 1.7$ for $n_\mathrm{m} = 0.1$ to $E = 10.1$ for $n_\mathrm{m} = 0.01$. Such an occupancy has been demonstrated on the same platform at lower pump
power in Ref.~\cite{werner2026electro}.

\begin{figure}[tb]
    \centering
\includegraphics[width=0.85\linewidth]{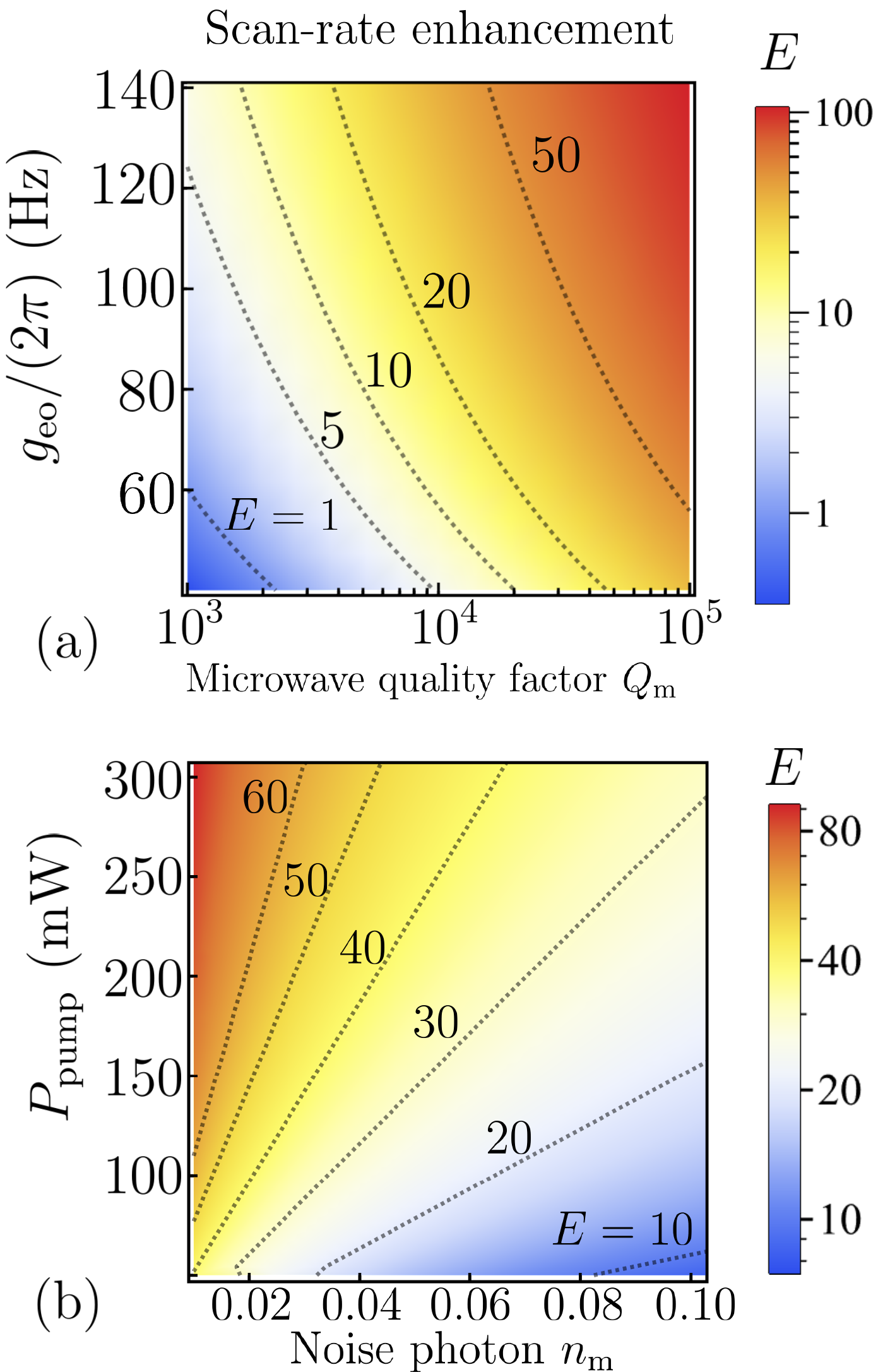}
    \caption{Scan-rate enhancement compared to standard homodyne detection under pulsed pumping. Dashed lines: the corresponding enhancement $E$. (a) Contour map of the optimal enhancement as a function of the $g_\mathrm{eo}$ and $Q_\mathrm{m}$, at a fixed noise photon $n_\mathrm{m}=0.01$ and pump power $P_\mathrm{pump}=100~ \mathrm{mW}$.  (b) Optimal enhancement versus $n_\mathrm{m}$ and $P_\mathrm{pump}$, at a fixed $Q_\mathrm{m}=5\times10^4$ and $g_\mathrm{eo}/(2\pi)=90~\mathrm{Hz}$.}
    \vspace{-3 mm}
    \label{fig:Pulse-enhancement}
\end{figure}

Within the quasi-steady-state approximation, the cooperativity $C$ can be expressed in terms of the single-photon coupling rate $g_\mathrm{eo}$, the pump power, and the intrinsic loss rates (i.e., the quality factors). Over the parameter ranges listed in Table~\ref{tab:parameters}, we evaluate the scan-rate enhancement. Fig.~\ref{fig:Pulse-enhancement}(a) maps the optimal enhancement over the $(g_\mathrm{eo}, Q_\mathrm{m})$ plane, while Fig.~\ref{fig:Pulse-enhancement}(b) shows it over the $(n_\mathrm{m}, P_\mathrm{pump})$ plane, with $(\beta_\mathrm{m}, \beta_\mathrm{o})$ optimized independently at each point.
The enhancement increases with a higher single-photon coupling rate $g_\mathrm{eo}$ and quality factor $Q_\mathrm{m}$, exceeding $E = 50$ in Fig.~\ref{fig:Pulse-enhancement}(a) when $g_\mathrm{eo}/(2\pi) = 90\,\mathrm{Hz}$ and $Q_\mathrm{m} = 5 \times 10^4$. Similarly, it exceeds $50$ in Fig.~\ref{fig:Pulse-enhancement}(b) when $n_\mathrm{m} \lesssim 0.02$ and $P_\mathrm{pump} \gtrsim 250\,\mathrm{mW}$.

\section{Discussion}
We have shown that converting axion-induced microwave photons to the optical
domain and counting them can substantially outperform conventional homodyne
readout. The scan-rate enhancement originates from that photon counting scheme avoids the
standard quantum limit of phase-preserving amplification. A central result of our analysis is that the optimal operating point is not the
one that maximizes the transduction efficiency. Because the figure of merit is
set by the balance between converted signal and transducer noise, the optimum
lies in the overcoupled regime, at a microwave coupling ratio $\beta_\mathrm{m}$
well above the value that maximizes $\eta_\mathrm{T}$. The efficiency is traded away
in exchange for a stronger suppression of the transducer noise. Current transduction experiments have not been developed for this specific application and operate in the near critical coupling regime to maximize efficiency. Still, the key ingredients for photon counting applications have been demonstrated on various platforms, both on-chip and 3D designs. The main challenge is to combine high cooperativity with low microwave occupancy, and optimal optical coupling to sustain this regime over long pump pulses.

Looking ahead, the scheme is attractive in high frequency,
where broadband quantum-limited phase-preserving
amplification remains challenging above $10$~GHz. The electro-optic interaction itself is not a
limitation in this range. The quantum transducers can be engineered to address microwave modes ranging from a few to ten GHz, provided that the microwave resonance matches the free spectral range of the optical cavity. The transducer bandwidth of
$1- 10~\mathrm{MHz}$ set by the microwave linewidth then fully covers
the signal linewidth. And the central conversion frequency can be tuned over
$100~\mathrm{MHz}$ by tuning the microwave cavity itself, allowing the
transducer to follow the haloscope cavity as it is stepped.


While our model is specific to electro-optic transduction, the framework may
be extended to other platforms, such as piezo-optomechanical
transducers, by including the corresponding noise and efficiency
contributions of the intermediate mechanical mode. 
Our proposed scheme could be further enhanced using two-mode squeezing input state and joint detection receiver as proposed in Refs. \cite{Shi2023,rosati2024joint,shi2025quantum, wang2026squeezing}.
Combining the readout with multi-cavity arrays or higher-order-mode cavities would allow the enhancement demonstrated here to multiplicatively compound with these architectures, translating into a massive boost in the scan rate and providing a concrete pathway toward high-speed axion searches at the $\mathrm{MHz/day}$ level in the 5--12~GHz range.

\section*{Acknowledgments}
This manuscript has been authored by the Fermi Forward Discovery Group, LLC under Contract No. 89243024CSC000002 with the U.S. Department of Energy, Office of Science, Office of High Energy Physics. The authors acknowledge support from the DOE Office of Science Early Career Research Program, Office of High Energy Physics (HEP), Detector R\&D and QIS programs, and the Fermilab Laboratory Directed R\&D (LDRD) program. R.C. acknowledges support from the U.S. Department of Energy, Office of
Science, National Quantum Information Science Research Centers, Superconducting Quantum Materials
and Systems Center (SQMS).

\section*{Data Availability}
The results presented in this work can be reproduced using the model and equations provided in the main text and Appendices, together with
the parameters listed in Table~\ref{tab:parameters}. The code used to generate the figures is available from the authors upon reasonable request.

\appendix
\section{Axion state in sensing cavity}
\label{apd:Axion-state}
The axion-induced field amplitude $b(\mathbf{r}, t)$ is expressed as
\begin{equation}
    b(\mathbf{r}, t) = \frac{1}{2} \left[ b(\mathbf{r}) e^{-i\omega_\mathrm{a} t} + b^*(\mathbf{r}) e^{i\omega_\mathrm{a} t} \right],
\end{equation}
where $\omega_\mathrm{a}$ is the eigenfrequency of the axion field. The coupling between the axion field and the electromagnetic field is described by the axion-photon interaction term in the Lagrangian
\begin{equation}
\mathcal{L}_{a\gamma\gamma} = -\frac{1}{4} g_{a\gamma\gamma} b(\mathbf{r}, t) F_{\mu\nu} \tilde{F}^{\mu\nu} = g_{a\gamma\gamma} b(\mathbf{r}, t) \mathbf{E} \cdot \mathbf{B},
\end{equation}
where $g_{a\gamma\gamma}$ is the axion-photon coupling constant, $F_{\mu\nu}$ is the electromagnetic field tensor, and $\tilde{F}^{\mu\nu}$ is its dual. Under the background magnetic field approximation where $\mathbf{B} \approx \mathbf{B}_0(\mathbf{r})$, keeping the first-order part of expression, the interaction Hamiltonian is given by
\begin{equation}
H_{\text{int}} \approx -\int_V g_{a\gamma\gamma} b(\mathbf{r}, t) \mathbf{B}_0(\mathbf{r}) \cdot \mathbf{E}(\mathbf{r}, t)\; d^3\mathbf{r}.
\end{equation}
The cavity mode expansion and the standard second quantization in natural units can be written as
\begin{align}
&\mathbf{E}(\mathbf{r},t) = i \sqrt{\frac{\omega_\mathrm{a}}{2}} (a e^{-i\omega_\mathrm{a}t} - a^\dagger e^{i\omega_\mathrm{a}t})\mathbf{e}(\mathbf{r}), \\ &b(\mathbf{r}, t) = \sqrt{\frac{1}{2 \omega_\mathrm{a} V_\mathrm{a}}} \left( \hat{b} e^{-i\omega_\mathrm{a}t} + \hat{b}^\dagger e^{i\omega_\mathrm{a}t} \right) \psi(\mathbf{r}),
\end{align}
where $\int \vert{}\psi(\mathbf{r})\vert{}^2 d^3\mathbf{r} = V_\mathrm{a}$ is the effective volume of the mode and $\int |\mathbf{e}(\mathbf{r})|^2 d^3\mathbf{r}=1$. Applying the rotating-wave approximation by neglecting the rapidly oscillating counter-rotating terms ($\hat{b}\hat{a}$ and $\hat{b}^\dagger\hat{a}^\dagger$), the interaction Hamiltonian can be simplified to
\begin{align}
H_{\text{int}} = \kappa_\mathrm{a} \left( \hat{b}^\dagger\hat{a} + \hat{b}\hat{a}^\dagger \right),
\end{align}
where the effective coupling strength $\kappa_\mathrm{a}$ is given by
\begin{align}
\kappa_\mathrm{a} = \frac{g_{a\gamma\gamma}}{2} \sqrt{\frac{1}{V_\mathrm{a}}} \int_V d^3\mathbf{r} \; \mathbf{B}_0(\mathbf{r}) \cdot \mathbf{e}(\mathbf{r}) \psi(\mathbf{r}).
\label{eq:kappa_a}
\end{align}
Here, $\kappa_\mathrm{a}$ quantifies the effective spatial overlap governing the axion-photon conversion. In natural units, taking typical values of $g_{a\gamma\gamma} \sim 10^{-15}\text{ GeV}^{-1}$ and a background magnetic field of $B_0 \approx 1.95 \times 10^{-15}\text{ GeV}^2$ (corresponding to $10\text{ T}$), the coupling rate evaluates to $\kappa_\mathrm{a} \sim 10^{-31}\text{ GeV}$ (approximately $10^{-6}\text{ s}^{-1}$). From $\frac{d\hat{a}}{dt} = -i[\hat{a}, H_{\text{int}}]$, the coupling between the axion field $\hat{b}$ and the cavity mode $\hat{a}$ is given by 
$
-i [\hat{a}, H_{\text{int}}] = -i\kappa_\mathrm{a} \hat{b}
$. Now, we consider a single-port (reflective) sensing cavity. The dynamics of the cavity mode $\hat{a}(t)$ is governed by the Heisenberg-Langevin equation in the time domain:
\begin{equation}
\frac{d\hat{a}}{dt} = -i \omega_\mathrm{c} \hat{a} - \frac{\kappa_{\mathrm{total}}}{2} \hat{a} - \sqrt{\kappa_{\mathrm{ext}}}\hat{a}_{\mathrm{in}} - \sqrt{\kappa_\mathrm{i}}\hat{a}_{\mathrm{i}} - i \kappa_a \hat{b},
\end{equation}
where $\omega_\mathrm{c}$ is the resonant frequency of the cavity, $\hat{a}_{\text{in}}$ and $\hat{a}_{\mathrm{i}}$ represent the input field and intrinsic noise operators, respectively, and $\kappa_{\mathrm{total}} = \kappa_{\mathrm{i}} + \kappa_{\mathrm{ext}}$, with $\kappa_{\mathrm{i}}$ and $\kappa_{\mathrm{ext}}$ denoting the intrinsic and extrinsic decay rates. To derive the steady-state transfer function, we transform the equations into the frequency domain using the Fourier convention $\hat{A}(\omega) = \frac{1}{\sqrt{2\pi}}\int_{-\infty}^{\infty} dt \, \hat{A}(t) e^{i\omega t}$. Hereafter, we work exclusively in the frequency domain unless stated otherwise. The resulting equation of motion and the corresponding input-output relation are given by:
\begin{align}
    -i\Delta\omega\hat{a} &= -\ \frac{\kappa_{\mathrm{total}}}{2}  \hat{a} - \sqrt{\kappa_{\mathrm{ext}}} \hat{a}_{\mathrm{in}}-\sqrt{\kappa_\mathrm{i}}\hat{a}_{\mathrm{i}}-i\kappa_a\hat{b},\\
    \hat{a}_{\mathrm{out}} &= \hat{a}_{\mathrm{in}} - \sqrt{\kappa_{\mathrm{ext}}} \hat{a},
\end{align}
where  $\Delta\omega=\omega-\omega_c$. From the above equations, the output mode in frequency domain is given by
\begin{align}
\hat{a}_{\mathrm{out}} &= C_{\mathrm{in}}(\omega) \hat{a}_{\mathrm{in}} + C_{\mathrm{i}}(\omega) \hat{a}_{\mathrm{i}} + C_{\mathrm{a}}(\omega) \hat{b},
\label{eq:input-ouput-sensing}
\end{align}
where the transfer coefficients are given by
\begin{subequations}
\begin{align}
C_{\mathrm{in}}(\omega) &= 1 - \frac{\kappa_{\mathrm{ext}}}{\frac{\kappa_{\mathrm{total}}}{2} - i\Delta\omega}, \\
C_{\mathrm{i}}(\omega) &= -\frac{\sqrt{\kappa_{\mathrm{ext}}\kappa_\mathrm{i}}}{\frac{\kappa_{\mathrm{total}}}{2} - i\Delta\omega}, \\
C_{\mathrm{a}}(\omega) &= - \frac{i\kappa_{\mathrm{a}}\sqrt{\kappa_{\mathrm{ext}}}}{\frac{\kappa_{\mathrm{total}}}{2} - i\Delta\omega}.
\end{align}
\end{subequations}
Treating the axion field as a classical driving source within the cavity, the resulting steady-state photon field forms a coherent state satisfying $\hat{b}\vert{}\alpha_\mathrm{a}\rangle = \alpha_\mathrm{a} \vert{}\alpha_\mathrm{a}\rangle$. The input noise modes are characterized by the vacuum state for $\hat{a}_{\mathrm{in}}$ and the thermal state for $\hat{a}_{\mathrm{i}}$ with the correlation function:
\begin{align}
\langle \hat{a}_{\mathrm{i}}^\dagger(\omega) \hat{a}_{\mathrm{i}}(\omega') \rangle = \delta(\omega - \omega') n_{\mathrm{s}}(\omega),
\end{align}
where the mean thermal photon number is given by the Bose-Einstein distribution
\begin{align}
n_{\mathrm{s}} = \frac{1}{e^{\hbar\omega/(k_BT_\mathrm{s})} - 1}.
\end{align}
In the absence of a thermal background, the power of axion signal can be derived from $P=\hbar \omega_\mathrm{a} \braket{a^\dagger_\mathrm{out}(t)a_{\mathrm{out}}(t)}\nonumber$. From $\hat{a}(t) = \int_{-\infty}^{\infty} \frac{d\omega}{\sqrt{2\pi}} \hat{a}(\omega) e^{-i\omega t}$, 
\begin{equation}
    \langle \hat{a}^\dagger(t) \hat{a}(t) \rangle = \int \int \frac{d\omega' d\omega}{2\pi} \langle \hat{a}^\dagger(\omega') \hat{a}(\omega) \rangle e^{i(\omega' - \omega)t}.
\end{equation}
For stationary process $\langle \hat{b}^\dagger(\omega') \hat{b}(\omega) \rangle = |\alpha_\mathrm{a}(\omega)|^2 \delta(\omega - \omega')$, where $\alpha_\mathrm{a}$ is the amplitude of coherent state. The power
\begin{align}
 P/(\hbar\omega_\mathrm{a})
 &=   \int_{-\infty}^\infty\braket{a^\dagger_\mathrm{out}(\omega)a_{\mathrm{out}}(\omega)} \;\frac{d \omega}{2\pi} \nonumber\\
 & = \int_{-\infty}^\infty|C_\mathrm{a}(\omega)|^2|\alpha_\mathrm{a}(\omega)|^2\;\frac{d \omega}{2\pi} \nonumber\\
 &\approx|C_\mathrm{a}(\omega_\mathrm{a})|^2|\alpha_0|^2\Delta f_\mathrm{a}.
 \label{eq:apd-power-1}
\end{align}
Here, we treat the axion signal $\alpha_\mathrm{a}(\omega)$ as a narrow-band process centered around $\omega_\mathrm{a}$ with an effective bandwidth $\Delta f_\mathrm{a}$ and define
\begin{align}
\int_{-\infty}^\infty |\alpha_\mathrm{a}(\omega)|^2 \; \frac{d\omega}{2\pi} =|\alpha_\mathrm{0}|^2 \Delta f_\mathrm{a},
\end{align}
where $|\alpha_0|^2$ represents the average spectral density.
In experiments, the relevant parameters are typically defined as
\begin{subequations}
\begin{align}
Q_L &\equiv  \frac{\omega_\mathrm{c}}{\kappa_{\mathrm{total}}}, \\
\frac{\beta}{\beta + 1} &\equiv  \frac{\kappa_{\mathrm{ext}} / \kappa_{\mathrm{i}}}{\kappa_{\mathrm{ext}} / \kappa_{\mathrm{i}} + 1} = \frac{\kappa_{\mathrm{ext}}}{\kappa_{\mathrm{total}}}, \\
\Delta &\equiv Q_L \frac{f_\mathrm{a} - f_\mathrm{c}}{f_\mathrm{c}}=\frac{f_\mathrm{a}-f_\mathrm{c}}{\Delta f_\mathrm{c}},
\end{align}
\end{subequations}
where $\beta = \kappa_{\mathrm{ext}}/\kappa_{\mathrm{i}}$ is the coupling coefficient, $Q_L$ is the loaded quality factor, and $\Delta$ represents the normalized detuning. Substituting these definitions into Eq.~\eqref{eq:apd-power-1}, we obtain the detected power:
\begin{align}
P = 4 \hbar Q_L \frac{\beta}{\beta + 1} \frac{1}{1 + 4\Delta^2} |\kappa_\mathrm{a}\alpha_0|^2 \Delta f_\mathrm{a}.
\label{eq:apd-power-2}
\end{align}
Comparing Eq.~\eqref{eq:apd-power-2} with the axion haloscope power formula,
\begin{align}
P = g_{a\gamma\gamma}^2 \frac{\rho_\mathrm{a}}{m_\mathrm{a}} B_0^2 V_{\mathrm{eff}} Q_L \frac{\beta}{\beta + 1} \frac{1}{1 + 4\Delta^2},
\end{align}
we establish the relation for the signal amplitude:
\begin{equation}
|\kappa_\mathrm{a}\alpha_\mathrm{0}|^2 = \frac{1}{4\hbar\Delta f_\mathrm{a}} g_{a\gamma\gamma}^2 \frac{\rho_\mathrm{a}}{m_\mathrm{a}} B_0^2 V_{\mathrm{eff}}.
\end{equation}
The real quadrature operators are defined as $\hat{q} = \frac{1}{\sqrt{2}}(\hat{a} + \hat{a}^\dagger)$ and $\hat{p} = \frac{i}{\sqrt{2}}(\hat{a}^\dagger - \hat{a})$, satisfying the commutation relation $[\hat{q}, \hat{p}] = i$.
For a coherent state $\vert{}\alpha_\mathrm{a}\rangle$, the corresponding displacement vector $\boldsymbol{d}$ (mean values) and the corresponding covariance matrix $\mathbf{V}$ can be obtained from above quadrature definitions:
\begin{align}
&\boldsymbol{d}_\mathrm{a} \equiv (\langle\hat{q}\rangle, \langle\hat{p}\rangle) = \sqrt{2} |\alpha_\mathrm{a}|(\cos\phi_\mathrm{a}, \sin\phi_\mathrm{a}), \\
&\mathbf{V}_\mathrm{a} \equiv \begin{pmatrix}
\langle\Delta\hat{q}^2\rangle & \frac{1}{2}\langle\{\Delta\hat{q}, \Delta\hat{p}\}\rangle \\
\frac{1}{2}\langle\{\Delta\hat{q}, \Delta\hat{p}\}\rangle & \langle\Delta\hat{p}^2\rangle
\end{pmatrix} = \frac{1}{2}\mathbf{I},
\end{align}
where $\alpha_\mathrm{a} = \vert{}\alpha_\mathrm{a}\vert{}e^{i\phi_\mathrm{a}}$, $\{\cdot, \cdot\}$ denotes the anti-commutator, and $\mathbf{I}$ is the $2 \times 2$ identity matrix. 
From Eq.~\eqref{eq:input-ouput-sensing}, we can derive the transformation of the mean vector and covariance matrix of the output state. Since $\boldsymbol{d}_\mathrm{in}=\boldsymbol{0}$, $\boldsymbol{d}_\mathrm{i}=0$, $\boldsymbol{V}_\mathrm{in} =\frac{1}{2} \boldsymbol{I}$ and $\boldsymbol{V}_\mathrm{i} =(\frac{1}{2}+n_\mathrm{s}) \boldsymbol{I}$, then
\begin{align}
    \boldsymbol{d}_{\mathrm{out}}
    &= \sqrt{2}|\alpha_\mathrm{a}(\omega)C_\mathrm{a}(\omega)|(\cos\phi_{\mathrm{a}},\sin\phi_\mathrm{a}),\\
    \boldsymbol{V}_{\mathrm{out}}&= \left(1/2+|C_\mathrm{i}|^2n_\mathrm{s}+|C_\mathrm{a}|^2/2\right) \boldsymbol{I}.
\end{align}
Since $\alpha_\mathrm{a}(\omega)$ is narrow-band process compared with the sensing cavity, $|\alpha_a(\omega)C_\mathrm{a}(\omega)|\approx|\alpha_0 C_\mathrm{a}(\omega_\mathrm{a})|$. By Eq.~\eqref{eq:apd-power-1},
\begin{align}
\boldsymbol{d}_{\mathrm{out}}&= \sqrt{2}\sqrt{\frac{P}{\hbar\omega_\mathrm{a} \Delta f_a}}(\cos\phi_{\mathrm{a}},\sin\phi_\mathrm{a})\nonumber\\
&\equiv\sqrt{2n_\mathrm{a}}(\cos\phi_{\mathrm{a}},\sin\phi_\mathrm{a}),
\label{eq:apd-mean}
\end{align}
where $n_\mathrm{a}=P/(\hbar\omega_\mathrm{a}\Delta f_\mathrm{a})$. Furthermore, since $\kappa_\mathrm{a}\approx10^{-6}\ll\sqrt{\kappa_\mathrm{i}}$, the output covariance matrix becomes
\begin{align}
\boldsymbol{V}_{\mathrm{out}}&= \left(1/2+|C_\mathrm{i}|^2n_\mathrm{s}\right) \boldsymbol{I} \nonumber\\
&=\left[1/2+\frac{4\beta}{(1+\beta)^2}\mathcal{L}(\Delta)n_\mathrm{s}\right] \boldsymbol{I},
\label{eq:apd-Vout}
\end{align}
where $\mathcal{L}(\Delta) \equiv 1/(1+4\Delta^2)$ is the Lorentzian shape function. 

\section{State transfer}
\subsection{Phase-Insensitive Amplifier and Noise Performance}
\label{sec:amplifier_noise}

The input-output relation for a general linear amplifier and a Josephson parametric amplifier (JPA) can be expressed as~\cite{PhysRevD.26.1817, RevModPhys.82.1155}
\begin{align}
    \hat{a}_{\mathrm{out}} &= \sqrt{G} \hat{a}_{\mathrm{in}} + \hat{F}, \\
    \hat{a}_{\mathrm{out}} &= \sqrt{G} \hat{a}_{\mathrm{in}} + \sqrt{G-1} \hat{a}_{\mathrm{e}}^\dagger,
\end{align}
where $[\hat{F},F^\dagger]=1-G$ and $\hat{a}_{\mathrm{e}}$ denotes the environmental annihilation operator. The thermal state of the environment is characterized by its mean thermal photon number,
\begin{equation}
    n_{\mathrm{JPA}} = \langle \hat{a}_\mathrm{e}^\dagger \hat{a}_\mathrm{e} \rangle = \frac{1}{e^{\hbar\omega/k_B T} - 1}.
\end{equation}
For a phase-insensitive broadband JPA, the covariance matrix of the environment state is given by $\boldsymbol{V}_{\mathrm{JPA}} = (1/2 + n_{\mathrm{JPA}})\bm{I}$. Consequently, the input-output mapping constitutes a Gaussian channel, yielding the output covariance matrix and displacement vector:
\begin{align}
    \boldsymbol{V}_{\mathrm{out}} &= G_{\mathrm{JPA}} \boldsymbol{V}_{\mathrm{in}} + (G_{\mathrm{JPA}} - 1) \boldsymbol{V}_{\mathrm{JPA}} \nonumber \label{eq:apd-Vout-JPA}\\
    &= \left[ G_{\mathrm{JPA}} \left(1 + |C_\mathrm{i}|^2 n_{\mathrm{s}} + n_{\mathrm{JPA}}\right) - \left(\frac{1}{2} + n_{\mathrm{JPA}}\right) \right] \bm{I}, \\
    \boldsymbol{d}_{\mathrm{out}} &= \sqrt{G_{\mathrm{JPA}}} \boldsymbol{d}_{\mathrm{in}}.
\end{align}
Defining the input-referred noise via the relation $\boldsymbol{V}_{\mathrm{out}} = \left(1/2 + G_{\mathrm{JPA}} n_{\mathrm{in,equiv}}\right) \bm{I}$, we obtain
\begin{align}
    n_{\mathrm{in,equiv}} &= 1 + |C_\mathrm{i}|^2 n_{\mathrm{s}} + n_{\mathrm{JPA}} - \frac{1 + n_{\mathrm{JPA}}}{G_{\mathrm{JPA}}} \nonumber \\
    &\approx 1 + |C_\mathrm{i}|^2 n_{\mathrm{s}} + n_{\mathrm{JPA}}.
\end{align}
To account for the subsequent amplification stages, we consider a concatenated high-electron-mobility transistor (HEMT) amplifier. The resulting output covariance matrix and displacement vector become
\begin{align}
    V_{\mathrm{out}} &= \left[ G_{\mathrm{HEMT}} G_{\mathrm{JPA}} \left(1 + |C_\mathrm{i}|^2 n_{\mathrm{s}} + n_{\mathrm{JPA}}\right) \right. \nonumber \\
    &\quad \left. + G_{\mathrm{HEMT}} (n_{\mathrm{HEMT}} - n_{\mathrm{JPA}}) - \left(\frac{1}{2} + n_{\mathrm{JPA}}\right) \right] \bm{I}, \\
    \boldsymbol{d}_{\mathrm{out}} &= \sqrt{G_{\mathrm{HEMT}} G_{\mathrm{JPA}}} \boldsymbol{d}_{\mathrm{in}}.
\end{align}
In the limit where $n_{\mathrm{HEMT}} \gg n_{\mathrm{JPA}}$, the input-referred noise simplifies to
\begin{equation}
    n_{\mathrm{in,equiv}} \approx 1 + |C_\mathrm{i}|^2 n_{\mathrm{s}} + n_{\mathrm{JPA}} + \frac{n_{\mathrm{HEMT}}}{G_{\mathrm{JPA}}}.
    \label{eq:apd-noise-transfer}
\end{equation}
By induction, the Friis formula for cascaded linear amplifiers, including the JPA, HEMT, and subsequent mixer, yields the comprehensive input-referred noise:
\begin{equation}
    n_{\mathrm{in,equiv}} = 1 + |C_\mathrm{i}|^2 n_{\mathrm{s}} + n_{\mathrm{JPA}} + \frac{n_{\mathrm{HEMT}}}{G_{\mathrm{JPA}}} + \frac{n_{\mathrm{mixer}}}{G_{\mathrm{JPA}} G_{\mathrm{HEMT}}}.
    \label{eq:apd-input-refered-noise}
\end{equation}
The total gain $G=G_\mathrm{JPA}G_\mathrm{HEMT}G_\mathrm{mixer}$.
Following this cascaded amplification chain, the final amplified signal state is characterized by the mean displacement vector and covariance matrix:
\begin{align}
    \boldsymbol{d}_{\mathrm{amp}} &= \sqrt{2Gn_\mathrm{a}}(\cos\phi_\mathrm{a}, \sin\phi_\mathrm{a}), \\
    \boldsymbol{V}_{\mathrm{amp}} &= \left(\frac{1}{2} + G n_{\mathrm{in,equiv}}\right) \bm{I}.
\end{align}
Finally, we note that the quantum limit $n_\mathrm{in,equiv}>1$ arises from two contributions in Eq.~\eqref{eq:apd-Vout-JPA}: the input quantum fluctuations $\boldsymbol{V}_\mathrm{in}$ and the amplifier fluctuations $\boldsymbol{V}_\mathrm{JPA}$. By employing a squeezed input and a corresponding phase-sensitive amplifier, one can achieve $\braket{\Delta q^2_\mathrm{in,equiv}}<1$ but $\braket{\Delta p^2_\mathrm{in,equiv}}>1$. This approach is known as squeezing-enhanced sensing.

\subsection{Quantum transduction}
\label{apd:transducer}

\begin{table}[htbp]
    \caption{Definitions of the parameters used in the transducer model.}
    \label{tab:variables}
    \begin{ruledtabular}
        \begin{tabular}{ll}
            \textbf{Symbol} & \textbf{Definition}\\
            \hline
            \multicolumn{2}{c}{\textit{Optical mode}} \\
            \hline
            $\omega_\mathrm{o,out}$ & Output optical frequency \\
$\gamma_\mathrm{oe}$ & Extrinsic coupling rate \\
$Q_\mathrm{o}$ & Intrinsic quality factor \\
$\gamma_\mathrm{oi}$ & Intrinsic loss rate, $\omega_\mathrm{o}/Q_\mathrm{o}$ \\
$\gamma_\mathrm{o}$ & Total damping rate, $\gamma_\mathrm{oe}+\gamma_\mathrm{oi}$ \\
$\Delta_\mathrm{o}$ & Detuning, $\omega_\mathrm{o,out}-\omega_\mathrm{o}$ \\
\hline
\multicolumn{2}{c}{\textit{Microwave mode}} \\
\hline
$\omega_\mathrm{m}$ & Microwave mode frequency \\
$\omega_\mathrm{m,out}$ & Output microwave frequency \\
$\gamma_\mathrm{me}$ & Extrinsic coupling rate \\
$Q_\mathrm{m}$ & Intrinsic quality factor \\
$\gamma_\mathrm{mi}$ & Intrinsic loss rate, $\omega_\mathrm{m}/Q_\mathrm{m}$ \\
$\gamma_\mathrm{m}$ & Total damping rate, $\gamma_\mathrm{me}+\gamma_\mathrm{mi}$ \\
$\Delta_\mathrm{m}$ & Detuning, $\omega_\mathrm{m,out}-\omega_\mathrm{m}$ \\
\hline
\multicolumn{2}{c}{\textit{Electro-optic coupling}} \\
\hline
$g_\mathrm{eo}$ & Single-photon coupling rate \\
$n_\mathrm{p}$ & Pump photon number \\
$g$ & Effective coupling strength, $\sqrt{n_\mathrm{p}}g_\mathrm{eo}$ 
            
        \end{tabular}
    \end{ruledtabular}
\end{table}

Assuming a steady-state conversion process, the input-output relations for electro-optical transduction in the frequency domain are given by:
\begin{subequations}
\begin{align}
    &0= i\Delta_\mathrm{o} \hat{a} - ig\hat{m} - \frac{\gamma_\mathrm{o}}{2}\hat{a} 
    + \sqrt{\gamma_{\mathrm{oe}}}\hat{a}_{\mathrm{in}} + \sqrt{\gamma_\mathrm{{oi}}}\hat{a}_{\mathrm{i}}, \\
    &0= i\Delta_\mathrm{m} \hat{m} - ig\hat{a} - \frac{\gamma_\mathrm{m}}{2}\hat{m} 
    + \sqrt{\gamma_{\mathrm{me}}}\hat{m}_{\mathrm{in}} + \sqrt{\gamma_\mathrm{{mi}}}\hat{m}_{\mathrm{i}}, \\
    &\hat{a}_{\mathrm{out}} = -\sqrt{\gamma_{\mathrm{oe}}}\hat{a} + \hat{a}_{\mathrm{in}},\\
    &\hat{m}_{\mathrm{out}}= -\sqrt{\gamma_{\mathrm{me}}}\hat{m} + \hat{m}_{\mathrm{in}},
\end{align}
\end{subequations}
where the variables and coefficients characterizing the transducer model are comprehensively summarized in Table~\ref{tab:variables}, categorized by the optical mode, microwave mode, and electro-optic coupling. Solving the quantum Langevin equations yields the optical output mode as
\begin{align}
    \hat{a}_{\mathrm{out}} &=  \left( 1 + \frac{\gamma_\mathrm{{oc}} X_\mathrm{m}}{X_\mathrm{o} X_\mathrm{m} + g^2} \right) \hat{a}_{\mathrm{in}}+\frac{\sqrt{\gamma_\mathrm{oe}}\sqrt{\gamma_\mathrm{oi}} X_\mathrm{m}}{X_o X_\mathrm{m} + g^2} \hat{a}_i \nonumber \\
    &+\frac{ig\sqrt{\gamma_{\mathrm{oe}}}\sqrt{\gamma_{\mathrm{me}}}}{X_\mathrm{o} X_\mathrm{m} + g^2} \hat{m}_{\mathrm{in}}+\frac{ig\sqrt{\gamma_{\mathrm{oe}}}\sqrt{\gamma_{mi}}}{X_\mathrm{o} X_\mathrm{m} + g^2} \hat{m}_\mathrm{i},
\end{align}
where $X_{\mathrm{o}} = i\Delta_{\mathrm{o}} - \frac{\gamma_{\mathrm{o}}}{2}$, and $X_{\mathrm{m}} = i\Delta_{\mathrm{m}} - \frac{\gamma_{\mathrm{m}}}{2}$. 
The conversion from $\hat{m}_{\mathrm{in}}$ to $\hat{a}_\mathrm{out}$ is determined by
\begin{equation}
\frac{ig\sqrt{\gamma_{\mathrm{oe}}}\sqrt{\gamma_{\mathrm{me}}}}{X_\mathrm{o} X_\mathrm{m} + g^2}  \hat{m}_{\mathrm{in}}.
\end{equation}
To maximize the conversion efficiency, the detunings are set to $\Delta_\mathrm{o} = \Delta_\mathrm{m} = 0$, yielding an efficiency of
\begin{equation}
    \eta_\mathrm{T} = \left\vert{} \frac{ig\sqrt{\gamma_\mathrm{oe}}\sqrt{\gamma_\mathrm{me}}}{\gamma_\mathrm{o}\gamma_\mathrm{m}/4 + g^2}\right\vert{}^2=\zeta_{\mathrm{m}} \zeta_{\mathrm{o}} \frac{4C}{(1+C)^2},
\end{equation}
where $C=4g^2/(\gamma_\mathrm{o} \gamma_\mathrm{m})$, $\zeta_\mathrm{o} = \frac{\gamma_\mathrm{oe}}{\gamma_\mathrm{o}}$ and  $\zeta_\mathrm{m} = \frac{\gamma_\mathrm{me}}{\gamma_\mathrm{m}}$.
Setting $\Delta_\mathrm{o}=0$ and $\Delta_\mathrm{m}=0$ respectively, the conversion bandwidths for the microwave and optical channels are given by
\begin{subequations}
    \begin{align}
        &\Delta\nu_{\text{FWHM}}(\Delta_{\mathrm{m}}) =\left\vert{} C -1  \right\vert{}\gamma_{\mathrm{m}},\\
        &\Delta\nu_{\text{FWHM}}(\Delta_{\mathrm{o}}) =\left\vert{} C -1  \right\vert{}\gamma_{\mathrm{o}}.
    \end{align}
\end{subequations}
Since these conversion bandwidths are typically greater than $1\text{ MHz}$—much larger than the typical signal bandwidth—the conversion efficiency can be treated as constant for the signal. 

At optical frequencies and room temperature, the thermal occupation of the optical input and intrinsic noise modes is negligible ($n_{\mathrm{o,in}}=n_{\mathrm{oi}} \approx 0$), whereas for the microwave noise mode, the correlation is given by
\begin{equation}
    \langle \hat{m}_{\mathrm{i}}^\dagger(\omega') \hat{m}_{\mathrm{i}}(\omega) \rangle = n_\mathrm{m} \delta(\omega' - \omega),
\end{equation}
where $n_\mathrm{m}$ represents the thermal photon number. The add noise photon from $\hat{m}_\mathrm{i}$ is given by
\begin{align}
     n_\mathrm{add}=&\left\vert{} \frac{ig\sqrt{\gamma_\mathrm{oe}}\sqrt{\gamma_\mathrm{mi}}}{\gamma_\mathrm{o}\gamma_\mathrm{m}/4 + g^2}\right\vert{}^2\int_{-\infty}^\infty d\omega\;\braket{m_\mathrm{i}^\dagger(\omega)m_\mathrm{i}(\omega)} \nonumber\\
     &=
     (1-\zeta_{\mathrm{m}}) \zeta_{\mathrm{o}} \frac{4C}{(1+C)^2} n_{\mathrm{m}}.
\end{align}
Given that the input microwave mode $\hat{m}_\mathrm{in}$ has the mean and covariance matrix specified in Eqs.~\eqref{eq:apd-mean} and \eqref{eq:apd-Vout}, the mean and covariance matrix of the output quantum state $\hat{a}_\mathrm{out}$ follow directly from their definitions and are given by:
\begin{align}
    \boldsymbol{d}_\mathrm{T}&= \sqrt{2\eta_\mathrm{T} n_\mathrm{a}}(\cos\phi_{\mathrm{a}},\sin\phi_\mathrm{a}),\\
    \boldsymbol{V}_\mathrm{T}&= \left(1/2+n_\mathrm{T}\right) \boldsymbol{I},
\end{align}
where the noise photon number is given by
\begin{align}
n_\mathrm{T}&=\zeta_{\mathrm{o}} \frac{4C}{(1+C)^2} \left[(1-\zeta_\mathrm{m})n_{\mathrm{m}}+\zeta_\mathrm{m}|C_\mathrm{i}(\Delta=0)|^2n_\mathrm{s}\right] \nonumber\\
&=\zeta_{\mathrm{o}} \frac{4C}{(1+C)^2} \left[(1-\zeta_\mathrm{m})n_{\mathrm{m}}+\zeta_\mathrm{m}\frac{4\beta}{(1+\beta)^2}n_\mathrm{s}\right].
\end{align}

\section{Signal-to-noise ratio and scan rate}
\subsection{Homodyne detection}
\label{apd:scan-rate-homodyne-detection}
In homodyne detection, a single frame of the power spectrum $q^2(\omega)$ is obtained via an analog-to-digital converter (ADC) and a fast Fourier transform (FFT). Each spectrum has a time duration of $1/\Delta\nu$, where $\Delta\nu$ is the spectral bin resolution. To suppress statistical noise, $N$ independent frames are collected, yielding a total integration time of
\begin{equation}
    t_{\mathrm{int}} = \frac{N}{\Delta\nu}.
\end{equation}
According to the central limit theorem, the sample mean of the power spectrum, $X = \frac{1}{N} \sum_{i=1}^{N} q_i^2$, approximately follows a normal distribution:
\begin{equation}
    X \sim \mathcal{N}\left(\mu = \langle \hat{q}^2 \rangle, \, \sigma^2 = \frac{\langle \hat{q}^4 \rangle - \langle \hat{q}^2 \rangle^2}{N}\right).
\end{equation}
When axion signal presents, the output state is a Gaussian state characterized by the mean displacement vector and covariance matrix:
\begin{align}
    \boldsymbol{d}_{\mathrm{amp}} &= \sqrt{2G n_\mathrm{a}} (\cos\phi_\mathrm{a}, \sin\phi_\mathrm{a}), \\
    \boldsymbol{V}_{\mathrm{amp}} &= \left(\frac{1}{2} + G n_{\mathrm{in,equiv}}\right) \bm{I}.
\end{align}
Assume that the phase $\phi_\mathrm{a}$ is uniformly distributed over the averaging interval, the moments of the quadratures are given by
\begin{align}
    &\langle \hat{q}^2 \rangle = \left(\frac{1}{2} + G n_{\text{in,equiv}}\right) + G n_\mathrm{a}, \\
    &\langle \hat{q}^4 \rangle - \langle \hat{q}^2 \rangle^2 = 4\left(\frac{1}{2} + G n_{\text{in,equiv}}\right) G n_\mathrm{a} \nonumber\\
    &\quad\quad\quad\quad\quad\quad+ 2\left(\frac{1}{2} + G n_{\text{in,equiv}}\right)^2.
\end{align}
Since $n_{\mathrm{in,equiv}} \gg n_\mathrm{a}$ and $G n_{\mathrm{in,equiv}} \gg 1$, the signal-to-noise ratio (SNR) simplifies to
\begin{equation}
    \text{SNR} = \frac{\mu_{\text{axion}} - \mu_{\text{no-axion}}}{\sigma_{\text{no-axion}}} \approx \frac{n_\mathrm{a}}{\sqrt{2} \, n_{\mathrm{in,equiv}}} \sqrt{t_\mathrm{int}\Delta\nu}.
\end{equation}
According to Eq.~\eqref{eq:apd-input-refered-noise}, the system's input-referred noise exhibits Lorentzian frequency dependence due to $|C_\mathrm{i}|^2n_\mathrm{s}$. Therefore, we can write the noise as
\begin{align}
    n_\mathrm{in,equiv}&= n_\mathrm{base}+(n_\mathrm{peak}-n_\mathrm{base})\mathcal{L}(\Delta), \label{eq:apd-in-equiv}
\end{align}
where $\mathcal{L}(\Delta)=1/(1+4\Delta^2)$ is the function of Lorentzian line shape and $n_\mathrm{base}>1$. 
Consider a scan process with a frequency step size of $\delta = \Delta f_\mathrm{scan}/N_2$, where $N_2$ is the total number of scanning steps. The total SNR is given by
\begin{align}
    \mathrm{SNR}^2_\mathrm{total} &= \sum_{k=1}^{N_2} \frac{n^2_\mathrm{a}(\Delta_k)}{2 n^2_{\mathrm{in,equiv}}(\Delta_k)} t_\mathrm{int}\Delta\nu \nonumber\\
    &= t_\mathrm{int}\Delta\nu\frac{1}{\delta}\sum_{k=1}^{N_2} \frac{n^2_\mathrm{a}(\Delta_k)}{2 n^2_{\mathrm{in,equiv}}(\Delta_k)} \delta\nonumber\\
    &\approx t_\mathrm{int}\Delta\nu \frac{1}{\delta} \int_{-\Delta f_\mathrm{scan}/2}^{\Delta f_\mathrm{scan}/2} df_\mathrm{a} \; \frac{n^2_\mathrm{a}(\Delta)}{2 n^2_{\mathrm{in,equiv}}(\Delta)} \nonumber\\
    &\approx t_\mathrm{int}\Delta\nu \frac{\Delta f_\mathrm{c}}{\delta} \int_{-\Delta f_\mathrm{scan}/(2\Delta f_\mathrm{c})}^{\Delta f_\mathrm{scan}/(2\Delta f_\mathrm{c})} d\Delta \; \frac{n^2_\mathrm{a}(\Delta)}{2 n^2_{\mathrm{in,equiv}}(\Delta)} \nonumber\\
    &\approx t_\mathrm{int}\Delta\nu \frac{\Delta f_\mathrm{c}}{\delta} \int_{-\infty}^{\infty} d\Delta \; \frac{n^2_\mathrm{a}(\Delta)}{2 n^2_{\mathrm{in,equiv}}(\Delta)}.
    \label{eq:apd-scan-rate-homodyne}
\end{align}
Here, we have converted the discrete summation into an integral approximation, which is valid for $\delta\ll\Delta f_\mathrm{c}$. And we assume a broad scan range such that $\Delta f_\mathrm{scan}/\Delta f_\mathrm{c} \gg 1$. Given that the scanning rate is defined as $R = \delta / t_\mathrm{int}$, the total squared SNR can be rewritten as
\begin{align}
    \mathrm{SNR}^2_\mathrm{total} = \frac{\Delta\nu \Delta f_\mathrm{c}}{R} \int_{-\infty}^{\infty} d\Delta \; \frac{n^2_\mathrm{a}(\Delta)}{2 n^2_{\mathrm{in,equiv}}(\Delta)}.
\end{align}
Substituting $n_\mathrm{a} = n_{\mathrm{a},0} \mathcal{L}(\Delta)$ and Eq.~\eqref{eq:apd-in-equiv}, we obtain
\begin{align}
    \mathrm{SNR}^2_\mathrm{total} = \frac{\pi}{8R} \frac{n_{\mathrm{a},0}^2}{n_\mathrm{peak}^{3/2} n_\mathrm{base}^{1/2}} \Delta f_\mathrm{c} \Delta\nu.
\end{align}

\subsection{Photon detection}
\label{apd:scan-rate-photon-detection}
A superconducting nanowire single-photon detector is not
photon-number resolving: it registers a click whenever one or more
photons are absorbed. The natural quantity is therefore the
complementary no-click probability. Since the signal absorption and the
dark-count process are statistically independent, then
\begin{align}
    P(\text{no click}) = P_{\mathrm{sig}}(0)\,P_{\mathrm{dark}}(0).
\end{align}
The signal is a Gaussian state with displacement
$\boldsymbol{d_{\mathrm{T}}}$ and covariance matrix
$\boldsymbol{V}_{\mathrm{T}}=(1/2+n_{\mathrm{T}})\boldsymbol{I}$. A detector of efficiency
$\eta_{\mathrm{D}}$ is modeled as a pure loss channel, which maps
$\boldsymbol{d_{\mathrm{T}}}\to\sqrt{\eta_{\mathrm{D}}}\,
\boldsymbol{d_{\mathrm{T}}}$ and
$n_{\mathrm{T}}\to\eta_{\mathrm{D}}n_{\mathrm{T}}$.
Then
\begin{align}
    P_{\mathrm{sig}}(0)=
    \frac{1}{1+\eta_{\mathrm{D}}n_{\mathrm{T}}}
      \exp\!\left(-\frac{\eta_{\mathrm{D}}
      |\boldsymbol{d_{\mathrm{T}}}|^{2}}
      {2(1+\eta_{\mathrm{D}}n_{\mathrm{T}})}\right),
\end{align}
Dark counts form an independent Poisson process with \begin{equation}
    P_\mathrm{dark}(k) = \frac{\lambda^k e^{-r_\mathrm{D}\tau }}{k!},
\end{equation} 
where $r_\mathrm{D}$ is the dark counting rate  and $\tau$ is the time window. A mode of bandwidth $\Delta f_{\mathrm{a}}$ is resolved
in a time window $\tau=1/\Delta f_{\mathrm{a}}$, so the number of dark counts per measurement window is
$n_{\mathrm{D}}=r_{\mathrm{D}}/\Delta f_{\mathrm{a}}$ and
$P_{\mathrm{dark}}(0)=e^{-n_{\mathrm{D}}}$. Combining these two factors,
\begin{align}
    P(\text{no click})
    = \frac{1}{1+\eta_{\mathrm{D}}n_{\mathrm{T}}}
     e^{-n_{\mathrm{D}}} \exp\!\left(-\frac{\eta_\mathrm{D}
      \eta_\mathrm{T} n_\mathrm{a}}
      {1+\eta_{\mathrm{D}}n_{\mathrm{T}}}\right)
      .
\end{align}
With $n_\mathrm{a}\ll n_\mathrm{T}$ and $n_\mathrm{D}\ll 1$, the probability of click is given by
\begin{align}
    \lambda\equiv P(\text{click}) \approx \frac{\eta_\mathrm{D} n_\mathrm{T}+n_\mathrm{D}}{1+\eta_\mathrm{D} n_\mathrm{T}} + \frac{\eta_\mathrm{D} \eta_\mathrm{T} n_\mathrm{a}}{(1+\eta_\mathrm{D} n_\mathrm{T})^2}. 
\end{align}
For a sufficiently long integration time $t_\mathrm{int}$, the number of independent modes is $N = t_\mathrm{int} f_\mathrm{r}$, where $f_\mathrm{r} = \Delta f_\mathrm{a}$ in the continuous-pump regime (set by the signal coherence bandwidth), and $f_\mathrm{r}$ is the pulse repetition rate in the pulsed-pump regime. The random variable $X=\sum_{i=1}^N \lambda_i/N$ has mean and variance
\begin{align}
    &\mu = \lambda,\\
    &\sigma^2 = \lambda(1-\lambda)/N.
\end{align}
The signal-to-noise ratio is given by
\begin{align}
    \text{SNR} &= \frac{\mu_\text{axion}-\mu_\text{no-axion}}{\sigma_{\text{no-axion}}}\approx\frac{\eta_\mathrm{D} \eta_\mathrm{T} n_\mathrm{a}}{\sqrt{ \eta_\mathrm{D} n_\mathrm{T}+n_\mathrm{D} }}\sqrt{t_\mathrm{int} f_r}.
\end{align}
Consider a scan process with a frequency step size of $\delta = \Delta f_\mathrm{scan}/N_2$, where $N_2$ is the total number of scanning steps. Similar to Eq.~\eqref{eq:apd-scan-rate-homodyne}, the total SNR is given by
\begin{align}
\mathrm{SNR}^2_\mathrm{total} &= \sum_{k=1}^{N_2} \frac{\eta_\mathrm{T}^2\eta_\mathrm{D}^2n^2_\mathrm{a}(\Delta_k)}{ \eta_\mathrm{D} n_\mathrm{T}+n_\mathrm{D}} t_\mathrm{int}f_\mathrm{r}\nonumber\\
&\approx  \frac{\eta_\mathrm{T}^2\eta_\mathrm{D}^2}{\eta_\mathrm{D} n_\mathrm{T}+n_\mathrm{D}} \frac{t_\mathrm{int}f_\mathrm{r}}{\delta}\int_{-\infty}^{\infty} df_\mathrm{a} \; n^2_\mathrm{a}(\Delta) \nonumber\\
&\approx\frac{\pi}{4R}\frac{\eta_\mathrm{D}^2\eta_\mathrm{T}^2n_\mathrm{a,0}^2}{\eta_\mathrm{D} n_\mathrm{T}+n_\mathrm{D}} \Delta f_\mathrm{c} f_\mathrm{r}.
\end{align}

\bibliographystyle{apsrev4-2}
\bibliography{Ref}

\end{document}